\documentclass[useAMS,usenatbib]{mn2e}

\usepackage{amsmath}
\usepackage{amssymb}
\usepackage{natbib}
\usepackage{graphicx}

\usepackage{multicol}

\newcommand{\jsource}{1RXS J180408.9-342058}
\newcommand{\exo}{EXO 1745-248}
\newcommand{\msource}{IGR J18245-2452}
\newcommand{\igr}{IGR J00291+5934}
\newcommand{\ngcsource}{Swift J0911.9-6452}

\usepackage[dvips]{color}

\def \aj {AJ}
\def \mnras {MNRAS}
\def \apj {ApJ}
\def \apjs {ApJS}

\def \apjl {ApJL}
\def \aap {A\&A}
\def \nat {Nature}
\def \araa {ARAA}

\def \pasj {PASJ}

\title[Rapid X-ray variability in very hard NS LMXBs]{Rapid X-ray variability properties during the unusual very hard state in neutron-star low-mass X-ray binaries}

\author[Wijnands et al.]{
\parbox[t]{\textwidth}{
\raggedright
R. Wijnands$^{1}$\thanks{r.a.d.wijnands@uva.nl}, A.S. Parikh$^{1}$, D. Altamirano$^{2}$, J. Homan$^{3}$, 
N. Degenaar$^{1}$
}
\vspace{6pt}\\
$^{1}$Anton Pannekoek Institute for Astronomy, 
University of Amsterdam,
Postbus 94249, 1090 GE Amsterdam, The Netherlands\\
$^{2}$Physics \& Astronomy, University of Southampton, Southampton, Hampshire SO17 1BJ, UK\\
$^{3}$MIT Kavli Institute for Astrophysics and Space Research, 77 Massachusetts Avenue 37-582D, Cambridge, MA 02139, USA
}

\begin{document}


\pagerange{\pageref{firstpage}--\pageref{lastpage}} \pubyear{0000}

\maketitle

\label{firstpage}

\begin{abstract}
Here we study the rapid X-ray variability (using {\it XMM-Newton} observations) of three neutron-star low-mass X-ray binaries (\jsource, \exo, and \msource) during their recently proposed very hard spectral state \citep[][]{parikh2017}. All our systems exhibit a strong to very strong noise component in their power density spectra (rms amplitudes ranging from 34\% to 102\%) with very low characteristic frequencies (as low as 0.01 Hz). These properties are more extreme than what is commonly observed in the canonical hard state of neutron-star low-mass X-ray binaries observed at X-ray luminosities similar to those we observe from our sources. This suggests that indeed the very hard state is a distinct spectral-timing state from the hard state, although we argue that the variability behaviour of {\msource} is very extreme and possibly this source was in a very unusual state. We also compare our results with the rapid X-ray variability of the accreting millisecond X-ray pulsars {\igr} and {\ngcsource} (also using {\it XMM-Newton} data) for which previously similar variability phenomena were observed. Although their energy spectra (as observed using the {\it Swift} X-ray telescope) were not necessarily as hard (i.e., for {\ngcsource}) as for our other three sources, we conclude that likely both sources were also in very similar state during their {\it XMM-Newton} observations. This suggest that different sources that are found in this new state might exhibit different spectral hardness and one has to study both the spectral as well as rapid variability to identify this unusual state.
\end{abstract}


\section{Introduction}

\begin{figure*}
 \begin{center}
\includegraphics[width=2\columnwidth]{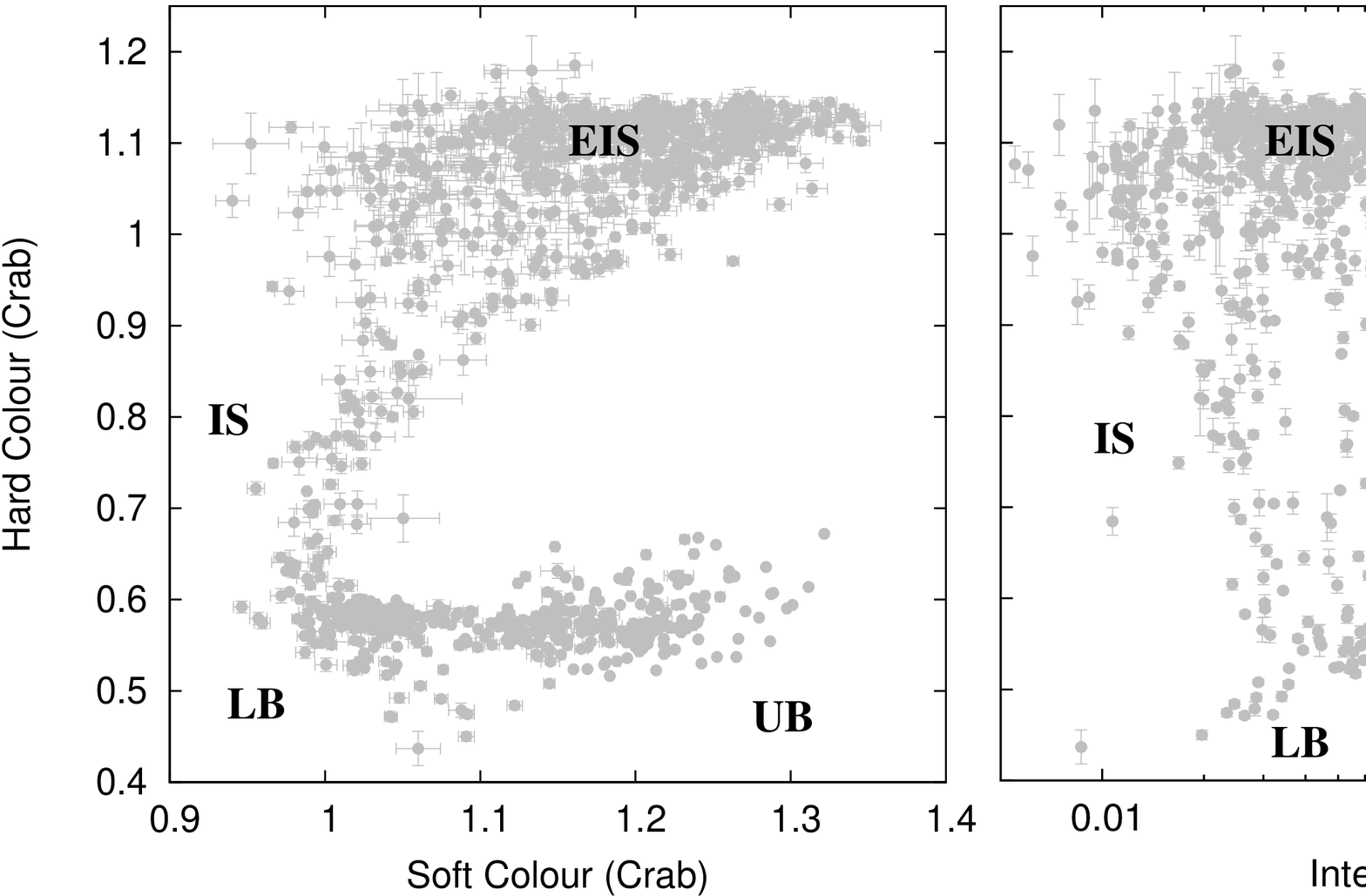}\
    \end{center}
 \caption[]{An example CD (left) and HID (right) for a typical atoll source (i.e., 4U 1608-52) that shows all states. The different branches/states are indicated in the figures (i.e., EIS = Extreme Island State; IS = Island State; LB = Lower Banana; UB = Upper Banana; the LB and the UB together form the banana branch). The figures were created using 1093 pointed observations on 4U 1608-52 with the proportional counter array (PCA) aboard  {\it RXTE} during the lifetime of this mission ($\sim16$ years). Each point represents the average colour/intensity per observation. Data were analysed in the exact same way as reported by \citet[][i.e., hard and soft colour are defined as the 9.7--16.0~keV~/~6.0--9.7~keV and 3.5--6.0~keV~/~2.0--3.5~keV count rate ratio, respectively, and intensity as the 2.0--16.0~keV count rate; note that the colours and intensities are normalised to the values observed for the Crab Nebula]{2008ApJ...685..436A}. We refer to that paper for the details about the reduction of PCA/{\it RXTE} data.} 
\label{fig:CDHID}
\end{figure*}

In neutron-star low-mass X-ray binaries (LMXBs), the compact star is accreting matter from a low-mass donor star (typically with masses $<1$ M$_\odot$). Traditionally, such systems are classified based on their correlated spectral and rapid variability behaviour. \citet[][]{1989A&A...225...79H} distinguished two main sub-types of neutron-star LMXBs: atoll sources and Z sources, named after the shape they trace out in their X-ray hardness-intensity diagram (HID) and colour-colour diagram \citep[CD; for more recent discussions see][]{2006csxs.book...39V,2010ApJ...719..201H}. Although it was clear for a long time that one of the main distinct differences between the two classes is that Z sources are accreting at much higher mass accretion rates (near Eddington rates) than atoll sources (few percent Eddington to at most several tens of percent Eddington), it was long debated if another property of the sources also contributed to the observed differences \citep[e.g., the magnetic field strength of the neutron stars; neutron-star spin rate; system inclination; for an in depth discussion see][]{1989A&A...225...79H}. Although hybrid behaviour was already seen in several sources \citep[e.g.,][]{1995A&A...297..141O,1999ApJ...522..965W,1999ApJ...512L..39W,2003A&A...406..221S,2004A&A...418..255H}, the breakthrough in our understanding came with the discovery of the bright X-ray transient XTE J1701-462 \citep[][]{2006ATel..696....1R}. By studying the huge amount of data obtained with the {\it Rossi X-ray Timing Explorer} ({\it RXTE}) taken during the outburst of this source, it became clear that the average mass accretion rate (averaged over days to weeks) solely determines the source class since this transient displayed Z source behaviour at the highest average luminosities but when this luminosity decayed it showed all characteristics of the atoll sources \citep[][]{2010ApJ...719..201H}. Related behaviour has also be seen in several additional luminous neutron-star LMXBs further strengthening this conclusion \citep[][]{2015ApJ...809...52F}.

For the current work, we will focus on the behaviour of neutron-star LMXBs when they have relatively low luminosities \citep[at most a few tens of percent Eddington, although a few systems might reach higher luminosities; see the large sample of sources displayed by][]{2007MNRAS.378...13G} and thus display atoll source characteristics. As is clear from the name, those sources traces out an atoll-shaped track in their CDs and HIDs (see Figure \ref{fig:CDHID} for typical examples of such diagrams for the well known atoll source and X-ray transient 4U 1608-52; see \citet[][]{2007MNRAS.378...13G} for additional examples). 

At the highest observed luminosities (and thus the highest inferred mass accretion rates), the track is called ‘banana branch’ which is often divided in ``upper'' banana (at the highest luminosities), and ``lower'' banana (at lower luminosities). The lower banana branch connects with the ``island'' state at the lowest banana-branch luminosities (this part is also called ``lower-left banana’’)\footnote{With sparse sampling of a particular source this connection is not always observed and often only the banana branch is observed with a small patch in the island state \citep[e.g.,][]{1989A&A...225...79H,2005ApJ...633..358A,2007MNRAS.378...13G}; this shape best resembles an atoll.}. In the island state the colours typically increase (the source becomes harder) when the inferred mass accretion rate decreases. However, the observed luminosity behaviour is complex.  In some sources the luminosity decreases as well, while in other sources it remains roughly the same and in some it might even increase again \citep[see Figure \ref{fig:CDHID} and the large source sample in][]{2007MNRAS.378...13G}. Some sources (i.e., the X-ray transients) show an extra patch at the largest hard colours called the ``extreme island'' state \citep[e.g.,][]{1997ApJ...485L..37M,2003ApJ...596.1155V}\footnote{We note that, to our knowledge, \citet[][]{1997ApJ...485L..37M} were  the first to use the term extreme island state in a paper. However, in that paper (as well in other papers in the late 1990’s) the term was used to indicate the extreme end of the island state \citep[e.g., in the brief review of][the extreme island state was not yet recognised as a separate state]{2001AdSpR..28..469W}. Only after the papers of \citet[][]{2002ApJ...568L..35M} and \citet[][]{2002MNRAS.331L..47G} it was realised that the extreme island state might constitute a different state (or branch in the CD and HID) of atoll sources (e.g., see the in depth discussions in \citet[][]{2003ApJ...596.1155V} and \citet[][]{2006csxs.book...39V}).}. This state has similar intensities\footnote{Again we would like to stress that the luminosity behaviour in the island and extreme island state is complex. The extreme island state can trace a large range of luminosities (e.g., see Figure \ref{fig:CDHID} right) and this range could overlap or even be larger than the luminosity range traced out during the island state. We also note that these are luminosities typically observed in a relatively narrow energy range and it is unclear if the bolometric luminosity behaves in the same way.} but smaller soft colours than the island state resulting in a horizontal extension in the CD and HID and therefore these sources form a Z shape as well \citep[][]{2002ApJ...568L..35M,2002MNRAS.331L..47G}. However, we note that the variability properties and detailed spectral behaviour observed when sources are in the extreme island state differ strongly from what is observed for Z sources when they are on the   horizontal branch of their Z track and therefore the extreme island state should not be confused with this Z-source horizontal branch \citep[e.g.,][]{2002ApJ...576..391B,2003ApJ...596.1155V}.

\begin{figure}
 \begin{center}
\includegraphics[width=\columnwidth]{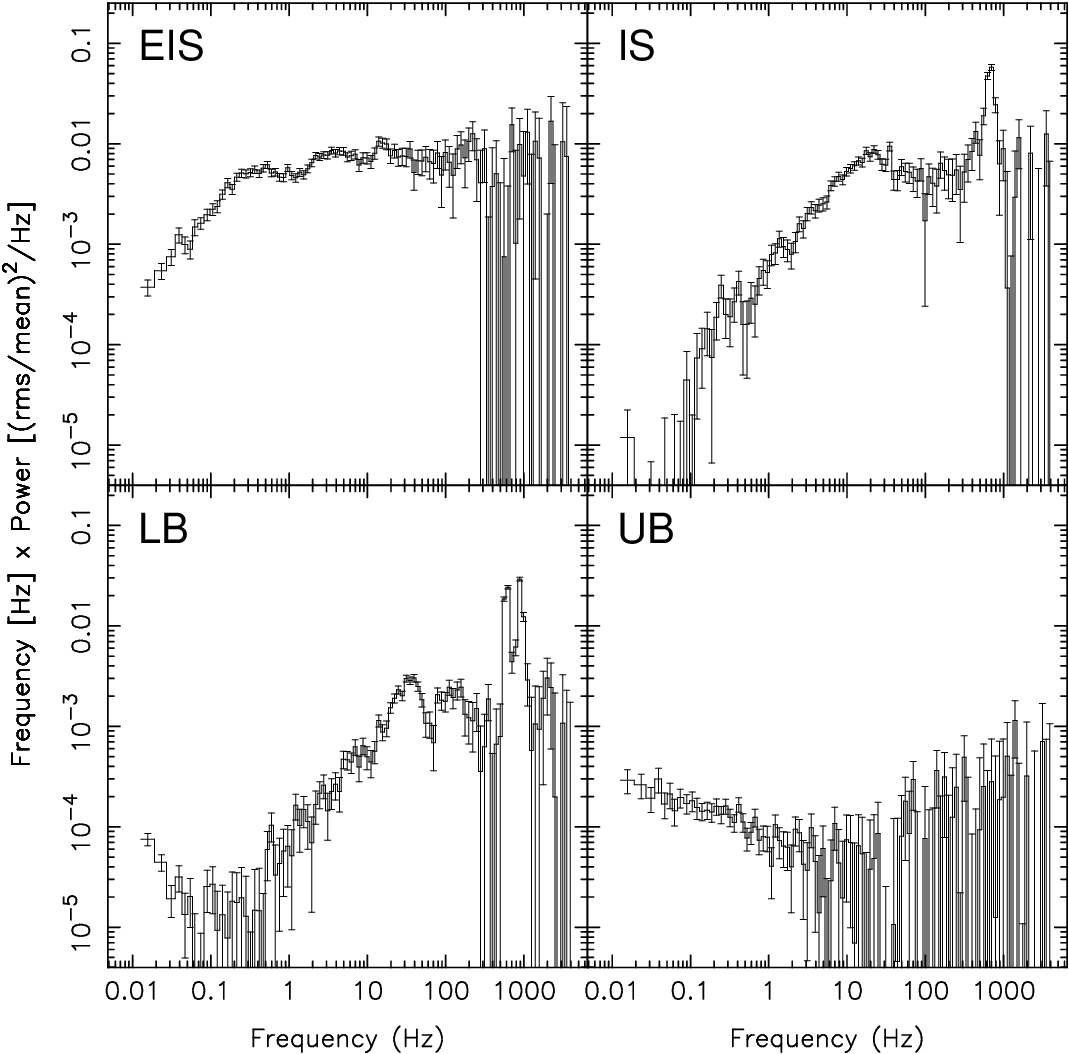}
    \end{center}
 \caption[]{Representative fractional-rms-normalized PDS created for the different atoll source states and branches (again for 4U 1608-52; made using PCA/{\it RXTE} data; see \citet[][]{2008ApJ...685..436A} for details on creating PDS from such data). The displayed PDS are created using individual observations drawn from the set of observations used to make Figure \ref{fig:CDHID}. In particular we used the following {\it RXTE} observations: observation ID numbers 30419-01-01-00 (EIS), 10094-01-02-00 (IS), 30062-02-01-01 (LLB), and 30188-01-02-00 (UB). } 
\label{fig:atollPDS}
\end{figure}

The rapid X-ray variability properties for atoll sources correlate very well with the position of the sources on the track \citep[e.g.,][]{1989A&A...225...79H,2002ApJ...568..912V,2003ApJ...596.1155V,2006csxs.book...39V,2008ApJ...685..436A}. In Figure \ref{fig:atollPDS} we show examples of the PDS typically observed during the different atoll source states and branches as described above \citep[for a similar figure see also Figure 3 of][]{2008ApJ...675.1407K}. In the extreme island state the sources have strong \citep[up to 30\%-40\% rms amplitude; e.g.,][]{1997ApJ...485L..37M,2002ApJ...568..912V} band-limited noise that has low characteristic frequencies (down to 0.1 Hz), visible as breaks in the noise and bumps (sometimes they are quasi-periodic oscillations, or QPOs) on top of the noise (typically at higher frequencies than the frequency at which the noise first breaks). This noise component decreases in strength (to $<$20\% rms) when the source moves from the extreme island state  to the island state and tends to disappear on the banana branch. During this evolution, the characteristic components increase in frequency and the bumps sometimes become more QPO like (if they were not already QPOs from the start; this depends on source). On the lower banana branch this band-limited noise is weaker (only a few percent rms amplitude) and can totally disappear on the upper banana branch. A weak (again a few percent rms) power-law shaped noise component (called the very-low frequency noise) becomes visible on the banana branch at the lowest probed frequencies. At the connection between the island state and the lower (left) banana also typically the kHz QPOs become visible in the PDS with frequencies between a few hundred Hz up to $\sim$1200 Hz. Those kHz QPOs usually come in pairs although depending on source and exact position of the source on the track only one of the QPOs might be visible \citep[for overviews see][]{2000ARA&A..38..717V,2006csxs.book...39V}

Very recently, surprisingly a potential new spectral state was identified at relatively high luminosities in three transient neutron-star LMXBs by \citet[][]{parikh2017}. When they fitted the energy spectra (in the energy range 0.5-10 keV) of a sample of sources with an absorbed power-law model, they found an unusual very hard state\footnote{\citet[][]{parikh2017} talked about soft state, hard state and very hard state, as is commonly done as well for neutron-star LMXBs (i.e., when also comparing neutron-star systems with black-hole ones). The soft state corresponds to the banana branch and the hard state corresponds to the island and extreme island state, with the extreme island state thought to be harder than the island state \citep[e.g.,][]{2003ApJ...596.1155V}.} between a luminosity of $10^{36}$ and $10^{37}$ erg s$^{-1}$. The hardness of the spectra (with photon indices of 1.2-1.4) obtained in this very hard state was significantly harder than the (extreme) island state spectra (with indices of 1.5-2.0) seen for atoll sources at the same luminosity. Although for individual sources these very hard spectra were already noted before \citep[][]{2014MNRAS.438..251L,2016MNRAS.460..345T},  \citet[][]{parikh2017} suggested that the sources were not in a very hard version of the (extreme) island state (as was assumed previously), but that the sources were in a new, previously unrecognised state. As explained above, the rapid variability properties of neutron-star LMXBs are very well correlated with spectral state. Therefore, in this paper we investigate the rapid variability of the three systems in their proposed very hard state to determine if indeed this state is significantly different from the extreme island state in other atoll sources.

\section{Observations, analysis and results} \label{sec:observations}

\citet[][]{parikh2017} identified three neutron-star LMXBs that exhibited very hard energy spectra: {\jsource}, {\exo}, and {\msource}. Each source exhibited an outburst in the last few years \citep[][]{2015ATel.6997....1K,2015ATel.7240....1A,2013ATel.4925....1E} and during those outbursts the sources were observed using {\it XMM-Newton}: once for {\exo}
\citep[][]{2017arXiv170307389M}, and twice for {\msource} and {\jsource}  \citep[][]{2013Natur.501..517P,2016ApJ...824...37L}. At least one {\it XMM-Newton} observation of each source was performed in its very hard state episode identified by \citet[][]{parikh2017}. Because of the brightness of the sources, all observations were performed using the EPIC-pn in its timing mode, resulting in a time resolution of $\sim$0.03 ms. This high time resolution (in combination with the high count rates) make these observations excellently suited to study the rapid X-ray variability behaviour of our targets during their very hard state. 

As we show below, the resulting power density spectra (PDS) show strong resemblances with those observed for the transiently accreting millisecond X-ray pulsar (AMXP) {\igr}  (during its 2004 outburst) by \citet[][]{2007ApJ...660..595L} using data obtained with {\it RXTE}. Since this source was also observed using {\it XMM-Newton} during its most recent outburst \cite[in 2015;][]{2017MNRAS.466.3450F} we included this source in our sample as well. Very recently, \citet[][]{2017ApJ...837...61B} suggested that the variability properties (studied using {\it XMM-Newton} and {\it NuSTAR} observations) of the transient AMXP {\ngcsource} \citep[during its 2016/2017 outburst\footnote{At the time of submitting our paper (end of July 2017), the source was still in outburst. };][]{2017A&A...598A..34S} resembled that of {\igr}. Therefore, we also included {\ngcsource} in our analysis. 

The full log of the observations used in our work is given in Table \ref{tab:log}. In Figures \ref{fig:1804_lcs}-\ref{fig:j0911_lcs}  we show, for each source in our sample, the outburst light curves and how the photon index (obtained from spectral fits in the 0.5-10 keV energy range) varies in time \cite[see also][]{parikh2017}. In those figures we have also indicated when the {\it XMM-Newton} observations were performed; it is clear that those observations were indeed performed during times when very hard spectra were observed for {\jsource}, {\exo} and {\msource} (bottom panels of Figures \ref{fig:1804_lcs}-\ref{fig:m28i_lcs}; photon indices $<$1.4). In addition,  {\igr} was also relatively hard (with a photon index of $\sim1.6$) at the times when its {\it XMM-Newton} observation was performed (bottom sub-panel of Figure \ref{fig:m28i_lcs} right). However, for {\ngcsource} the index is on average $\sim$1.7-1.8 (bottom panel of Figure \ref{fig:j0911_lcs}) which is similar to what is observed for other neutron-star LMXBs when they have similar luminosities as we observed for {\ngcsource}. Thus this source is not unusually hard (see also Figure \ref{fig:gammavsLx}). In Section \ref{subsec:igr} we discuss in more detail how this source compares to our other sources.

\begin{figure*}
 \begin{center}
\includegraphics[width=1.398\columnwidth]{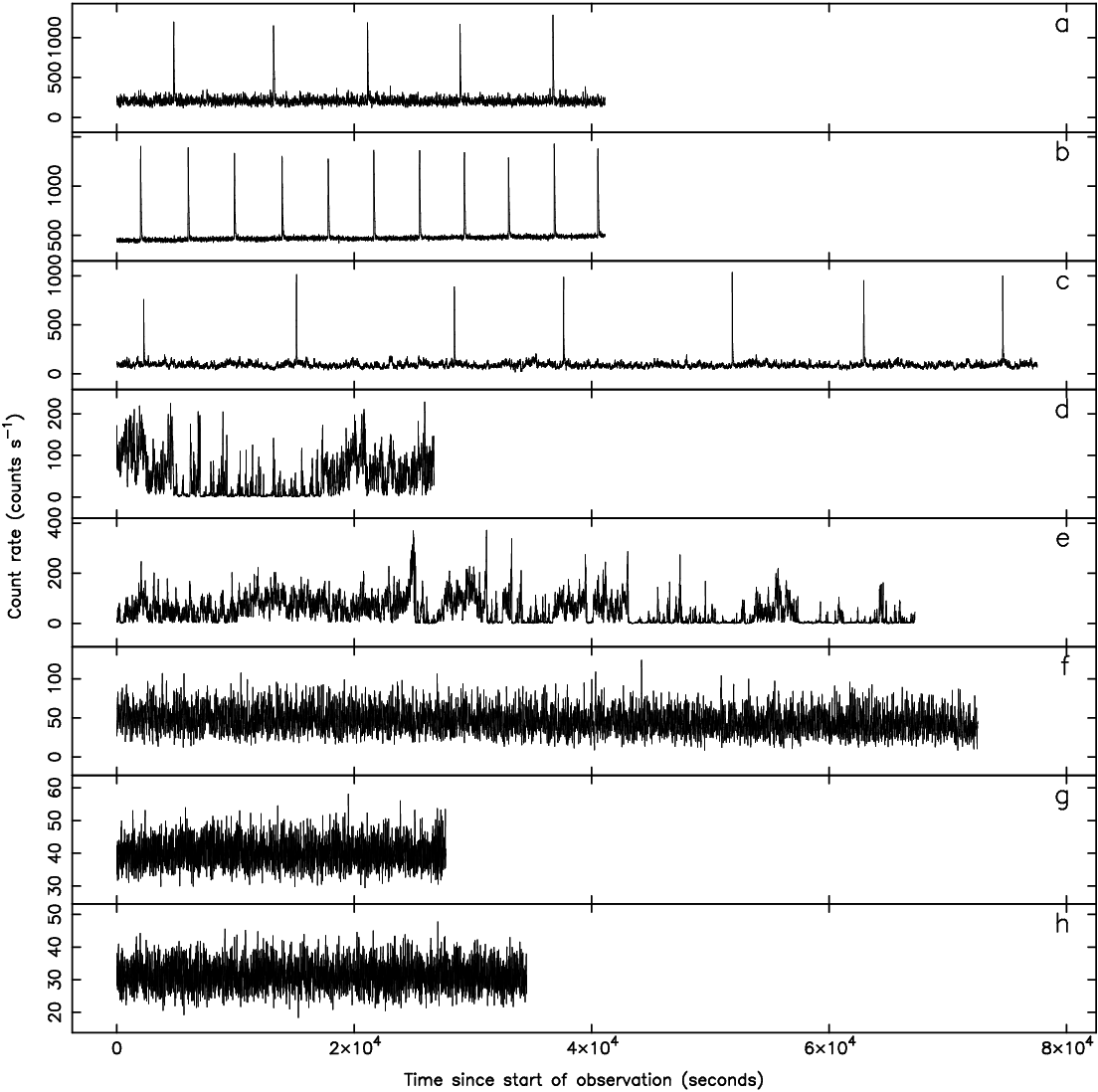}\hspace{0.75cm}
\includegraphics[width=0.55\columnwidth]{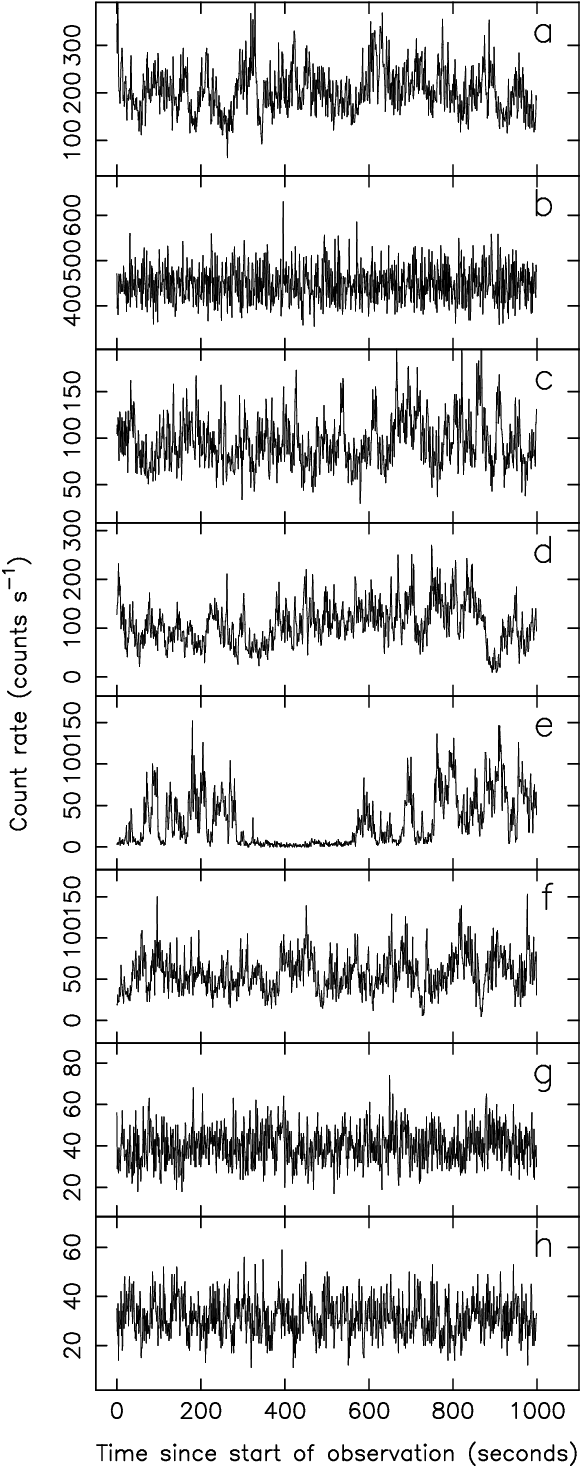}
    \end{center}
 \caption[]{The 0.5-10 keV light curves (10 seconds resolution) of our sample of sources: (a and b) {\jsource} during its first and second observation (respectively), (c) {\exo}, (d and e) {\msource} during its first and second  observation (respectively), (f) {\igr}, and (g and h) {\ngcsource} during its first and second observation (respectively). Left: the light curves for the whole observation; right: the light curves for the first thousand seconds of each observation.} 
\label{fig:10secondLCs}
\end{figure*}

All data were obtained from the {\it XMM-Newton} Science Archive\footnote{https://www.cosmos.esa.int/web/xmm-newton/xsa} and the data were processed and reduced using the {\it XMM-Newton} Scientific Analysis System (SAS\footnote{https://www.cosmos.esa.int/web/xmm-newton/sas}; version 15.0.0). When a source is observed with {\it XMM-Newton}, data are collected simultaneously  using all detectors on board the satellite. However, here we will only use the EPIC-pn data, since, as stated above, these data allowed us to study the X-ray variability properties of our sources up to high frequencies. To produce pn events list that incorporate the most up to date calibration information, we used the tool {\em epproc}. 

The pn in timing mode allows both timing analysis and spectral analysis in the energy range 0.3-10 keV. However, in this paper we are only interested in the variability properties of our sources. Therefore we do not analyse the pn spectra in our paper \citep[for detailed spectral analyses of the {\it XMM-Newton} data for our sources we refer to][]{2014A&A...567A..77F,2016ApJ...824...37L,2017A&A...598A..34S,2017MNRAS.466.2910S,2017arXiv170307389M,2017arXiv170404181D}.

\begin{table}
\caption{Log of the {\itshape XMM-Newton} observations}
\begin{tabular}{llcc}
\hline
Source                & ObsID         & Start of exposure &  Time$^a$ \\
                      &               &  (UTC)            &   (ksec) \\
\hline 
\multicolumn{4}{c} {\underline {Sources identified as very hard by \citet{parikh2017}}}\\
{\jsource}            &  0741620101   & 2015-03-06 18:38 & 41.1 \\
                      &  0741620201   & 2015-04-01 17:29 & 40.9 \\
{\exo}                &  0744170201   & 2015-03-23 05:44 & 77.5 \\
{\msource}            &  0701981401   & 2013-04-04 00:24 & 26.7 \\
                      &  0701981501   & 2013-04-13 07:00 & 67.2 \\
\hline
\multicolumn{4}{c}{\underline{Additional sources studied in the current work}}\\
{\igr}                &  0744840201   & 2015-07-28 12:40 & 72.5 \\
{\ngcsource}          &  0790181401   & 2016-04-24 05:20 & 27.7\\
                      &  0790181501   & 2016-05-22 14:15 & 34.5\\
\hline
\multicolumn{4}{l} {$^a$ Exposure time of the EPIC-pn detector }
\end{tabular}
\label{tab:log}
\end{table}

\subsection{Type-I X-ray bursts}\label{sec:bursts}

The light curves of our sources (10 seconds time resolution; for the energy range 0.5-10 keV; obtained using the same source extraction regions as used in Section \ref{sec:timing}) are shown in Figure \ref{fig:10secondLCs}. Type-I X-ray bursts are clearly visible for {\jsource} and {\exo} but no bursts were observed for the other three sources. 

During the first observation of {\jsource} five type-I X-ray bursts were seen in $\sim$40.9 ksec giving a burst rate of one bursts per $\sim$8.2 ksec (or one per $\sim$2.3 hours). \citet[][]{2016ApJ...824...37L} stated that they observed seven type-I bursts during this {\it XMM-Newton} observation but they also analysed $\sim9.5$ ksec of data during which the pn was in burst mode (in addition to the pn timing mode data). Those burst mode data are not used in our paper but we checked and indeed the two extra bursts were clearly visible in those data.  During the second observation of this source, eleven bursts were seen in $\sim$41.1 ksec resulting in one burst every $\sim$3.7 ksec (or approximately one burst per hour). The persistent count rate (0.5-10 keV) increased from $\sim$200 count s$^{-1}$ during the first observation to $\sim$470 counts s$^{-1}$ during the second observation. The factor 2.2 increase in burst frequency between the two observations roughly matches the factor 2.4 increase in persistent brightness. Assuming that the burst rate increases proportionally with the mass accretion rate \citep[see, e.g.,][for discussions and the problems with this hypothesis]{1981ApJ...247..267F,1998ASIC..515..419B,2008ApJS..179..360G} this would indicate that the accretion rate onto the neutron-star surface increased by a factor of $\sim$2 as well between the two observations.

EXO 1745-248 exhibited seven type-I bursts in $\sim$77.5 ksec \citep[for a detailed study of the bursts in this source see][]{2017arXiv170307389M}, resulting in one burst every $\sim$11.1 ksec (one burst every $\sim$3.1 hour), although the burst rate was not fully constant during the observation. For example, the third burst occurred $\sim$13.3 ksec after the second one, but then the fourth burst already occurred $\sim$9.2 ksec after the third one (although the averaged recurrence time of $\sim$11.3 ksec between those three bursts is very close to the recurrence time when all bursts are considered).

\begin{figure*}
 \begin{center}
\includegraphics[width=\columnwidth]{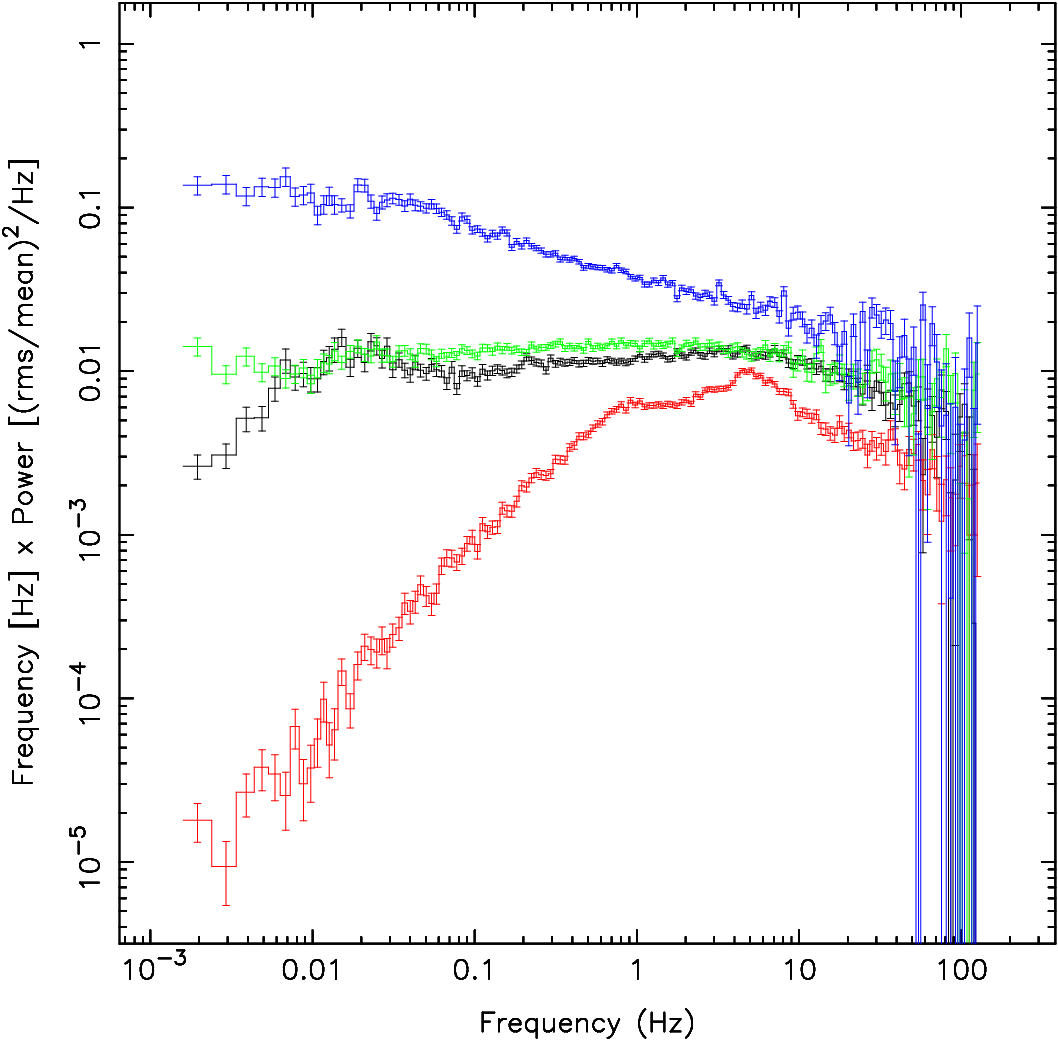} \hspace{0.5cm}
\includegraphics[width=\columnwidth]{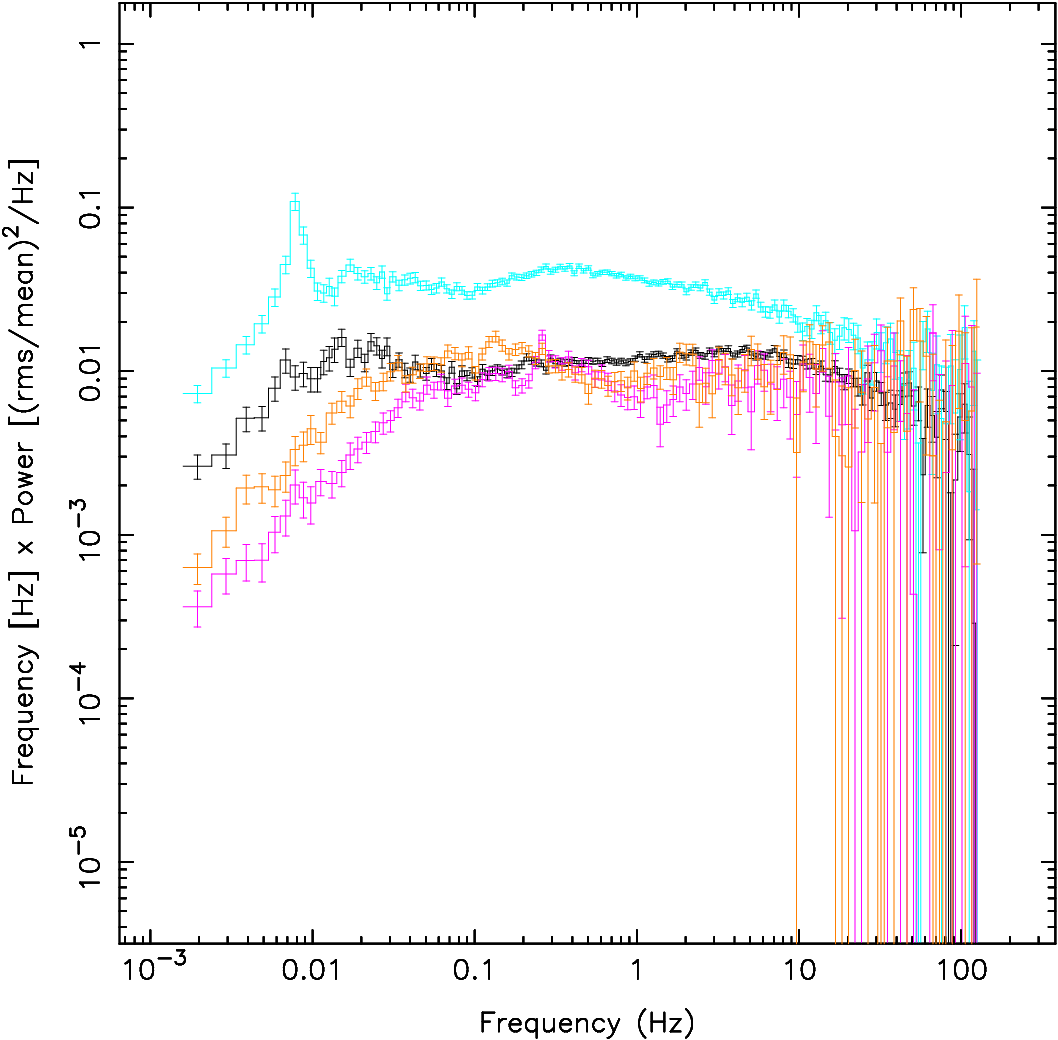}
    \end{center}
 \caption[]{{\it Left}: The PDS of {\jsource} (black: first observation; red: second observation), {\exo} (green), and {\msource} (dark blue; second observation of the source). {\it Right}: The PDS of {\igr} (light blue) and {\ngcsource} (pink: first observation; brown: second observation). Also for comparison we show the PDS obtained during the first observation of {\jsource} (black).}
\label{fig:allPDS}
\end{figure*}

\subsection{X-ray timing analysis}\label{sec:timing}

Since the EPIC-pn detector was used in timing mode, one of the spatial dimensions was sacrificed to obtain the high time resolution \citep[][]{2001A&A...365L..18S}. Therefore, the source extraction regions had to be defined in the RAWX and RAWY coordinates. However, due to the brightness of our sources and the fact that the size of the read-out area of the EPIC-pn in timing mode is smaller than the point-spread-function of the telescope, the source photons were spread out over the whole read-out area \citep[at least for part of the energy range; e.g., see discussion in][]{2010A&A...522A..96N}. Consequently, the optimal source extraction region is not trivial to choose because it depends on determining the best signal-to-noise (SNR) of the resulting data which depends on the size of the extraction region and on the brightness of the source. Another complication is that to renormalise the resulting PDS to squared fractional rms also the background count rate needs to be obtained. However, the standard way of extracting those rates from a source free region on the CCD cannot be applied since such a region does not exist in timing mode for sources as bright as our targets. 

In Appendix \ref{sec:timingtesting} we investigated the effect of using different extraction regions on the signal-to-noise ratio of the resulting PDS. We also investigated the systematic uncertainties on the rms amplitude of the broad-band noise due to the difficulties in determining the background count rates, as well as several other effects (e.g., pile-up) that might hamper us in obtaining the correct rms amplitude.  We found that the best source extraction region depends strongly on the source brightness. To have a homogenous analysis among our objects we settled on a source extraction region of 20 pixels (centred on the source position; determined as the column in the RAWX-RAWY image where the number of counts was highest). This is a compromise between being able to optimally constrain the PDS and minimising the effect of the background uncertainties (for which we decided not to correct for when we renormalised our PDS; see Appendix \ref{sec:timingtesting} for full details).

To broadly investigate the energy dependency of the PDS, we extracted the source event lists in several energy ranges: 0.5-10 keV, 0.5-1 keV, 1-2 keV, and 2-10 keV. For {\exo} we also created event list in the 1-10 keV energy range (see Section \ref{subsection:exo}). We rebinned the data to a time resolution of $2^{-11}$ seconds (resulting in a Nyquist frequency of 1024 Hz) and calculated fast Fourier transforms (FFTs) of the data to create PDS using a variety of lengths for the data segments. Preferably, we would like to make PDS with the length of the duration of the full observations to study the variability behaviour up to the lowest possible frequencies. However, {\jsource} and {\exo} exhibited many type-I X-ray bursts (Section \ref{sec:bursts}; Fig.~\ref{fig:10secondLCs}). Those bursts had to be removed from the data before PDS could be created (when using FFTs) and the maximum useable length for the PDS was constrained by the shortest recurrence time of the bursts, i.e., during the second observation of {\jsource}.  The average time interval between bursts in this observation was $\sim$3.7 ksec (see Section \ref{sec:bursts}). This duration would allow for PDS to be created that have a length of 1024 or 2048 seconds. However, when creating PDS using the latter length, only 2048 seconds of each interval between the bursts is used in our final analysis and thus a significant amount of data ($\sim$1700 second per interval between bursts) is thrown away. Therefore, we settled on making PDS with a length of 1024 seconds (giving a minimum frequency we can probe of 1/1024 Hz; we now only do not use $\sim$600 seconds of data in between the type-I bursts for this observation). For a homogenous analysis we used the same length to create the PDS for  all our sources.

When creating the PDS, no background or dead-time corrections were made. All PDS were combined per observation to create one averaged power spectrum.  The Poisson level was assumed to be constant and was estimated between 350 and 450 Hz where no noise is expected \citep[and neither any contributions from the pulsations for the AMXPs in our sample;][]{2005ApJ...622L..45G,2013Natur.501..517P,2017A&A...598A..34S} and where none of the instrumental spikes above $\sim$160 Hz \citep[called the critical frequency by][]{1999SPIE.3765..673K,2002astro.ph..3207K} are present\footnote{We have investigated the effect of using different frequency intervals  (always excluding the instrumental spikes or pulsations) to estimate the Poisson level, but we found no significant effect on the resulting PDS.}. The obtained estimate for the Poisson level was subtracted from the averaged PDS. The PDS were converted to squared fractional rms using a zero background count rate (see Appendix \ref{sec:timingtesting} for the justification of this choice). Therefore the rms amplitudes quoted in this paper are formally only lower limits to the actual values but should be close to the real ones. 

\begin{figure}
 \begin{center}
\includegraphics[width=\columnwidth]{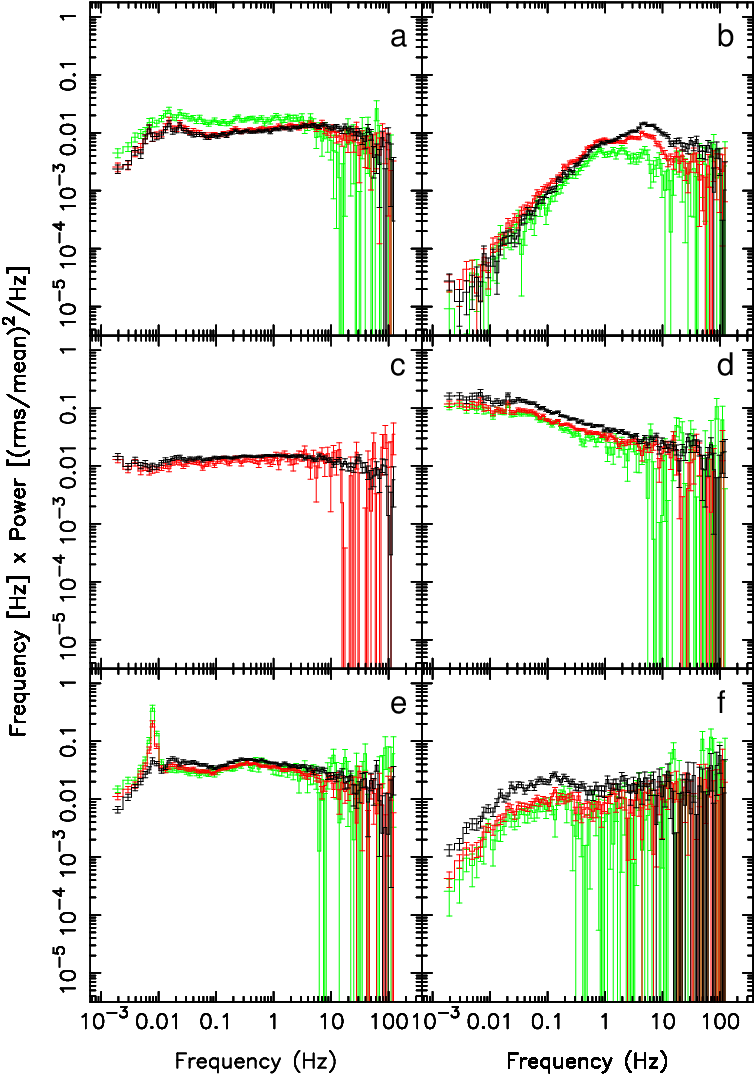}\
    \end{center}
 \caption[]{The PDS of our sample of sources in different energy ranges (green: 0.5-1 keV; red: 1-2 keV; black: 2-10 keV). In panel a we show the PDS for {\jsource} during its first {\it XMM-Newton} observation, in panel b the ones of {\jsource} during its second observation; in panel c the ones of {\exo} (we do not show the 0.5-1 keV power spectrum here because of its very low quality; see Section \ref{subsection:exo}), in panel d the ones of {\msource} (second observation only), in panel e the ones of {\igr}, and in panel f the ones of {\ngcsource} (second observation only). }
\label{fig:energydependency}
\end{figure}

The PDS are plotted in our figures in the frequency times power ($\nu P_\nu$) representation \citep[][]{1997A&A...322..857B}. The PDS of {\jsource}, {\exo}, and {\msource} are shown in Figure \ref{fig:allPDS}, left panel. The PDS of  {\igr} and {\ngcsource} are shown in Figure \ref{fig:allPDS}, right panel, together again (for comparison) with the one of {\jsource} obtained during its first {\it XMM-Newton} observation (i.e., in its very hard state).  We only show the PDS up to 128 Hz to avoid displaying the instrumental spikes occurring above $\sim$160 Hz \citep[see, e.g.,][]{1999SPIE.3765..673K,2002astro.ph..3207K}. For each observation, we also calculated PDS for the different energy ranges we selected and the results are shown in Figure \ref{fig:energydependency}.

Typically the PDS of neutron-star LMXBs are fitted with a set of Lorentzian functions and then the different Lorentzians are identified to compare their properties between various sources \citep[e.g.,][]{2002ApJ...568..912V,2003ApJ...596.1155V,2008ApJ...685..436A}. However, despite the fact that the shape of the PDS of the individual sources resemble each other (except the one of \msource), the detailed shape is quite different, making a detailed comparison complex and uncertain. In addition, {\msource} showed quite a different PDS (Section \ref{subsec:m28i}) and a large number of Lorentzians need to be used to fit the PDS adequately. In the current work we are only interested in the broad picture and we defer detailed analyses of our PDS to future work (but see already \citet[][]{2017MNRAS.466.3450F} and \citet[][]{2017ApJ...837...61B} for such detailed studies for {\igr} and {\ngcsource}, respectively). To determine the strength of the noise components we observed in the PDS, we calculated the integrated noise strengths between $9.765625 \times 10^{-4}$ and 128 Hz ($2^{-10}$ - $2^7$ Hz). In the next few subsections we will discuss the general behaviour of each source separately.

\subsection{\jsource}

The two PDS created for {\jsource} during the two {\it XMM-Newton} observations differ markedly. In the PDS of the first observation (black curve in Figure \ref{fig:allPDS}) the source exhibited strong band limited noise that is spread over the full displayed frequency range (0.001-100 Hz). The noise component peaks in strength at a relatively low characteristic frequency of $\sim$0.01 Hz and has a second peak at around 10 Hz. However, the strength of the noise component only minorly decreases between those two frequencies. At frequencies below 0.01 Hz and above 10 Hz, the noise decreases significantly in strength. The noise has an integrated rms value of 33.5$\pm$0.2\% (0.5-10 keV). It was strongest in the lowest energy band (0.5-1 keV); in the two highest energy bands the noise strengths were very similar to each other (see Figure \ref{fig:energydependency}, panel a, and Table \ref{tab:energydependency}). However, at the highest frequencies ($>$10 Hz) the noise is actually weakest in the lowest energy band, demonstrating that the shape of the PDS changes significantly with photon energy (Figure \ref{fig:energydependency}, panel a).

The PDS created for the second {\it XMM-Newton} observation (red curve in left panel of Figure \ref{fig:allPDS}) has a much more peaked shape: the noise peaks at $\sim$1 Hz with a bump on top of the noise at $\sim$5 Hz. This shape resembles the canonical island state PDS observed for atoll sources \citep[e.g., see Figure \ref{fig:atollPDS}; see also the extensive studies by][]{2002ApJ...568..912V,2003ApJ...596.1155V,2005ApJ...633..358A,2008ApJ...685..436A}. A correlation has been found between the  characteristics frequencies of the main noise component and the bump that is roughly the same for many sources \citep[][]{1999ApJ...514..939W}. The PDS of {\jsource} during the second observation is also consistent\footnote{When this PDS  of {\jsource} is fitted with a broken power-law model plus a Lorentzian, the frequencies of those components are 0.72$^{+0.03}_{-0.01}$ Hz and 4.1$\pm$0.1 Hz, respectively. These values are fully consistent with the relation found by \citet[][]{1999ApJ...514..939W} for these components.} with this correlation confirming the identification of this PDS with the island state PDS. The integrate noise only has an rms amplitude of 18.2$\pm$0.1\% (0.5-10 keV), which is significantly lower than what we found during  the first {\it XMM-Newton} observation. Contrary to what we found for the PDS obtained during the first observation, for the second observation, the integrated noise in the PDS increases gradually in strength with increasing photon energy, from $\sim$14\% (0.5-1 keV) to $\sim$22\% (2-10 keV; Table \ref{tab:energydependency}). But as can be seen from Figure \ref{fig:energydependency} (panel b) the full energy dependency is more complex: the noise increased in strength for all frequencies when the photon energies increase from 0.5-1 keV to 1-2 keV, however, for higher energies  the strength decreases again at the low frequencies ($<1$ Hz) while for higher frequencies the noise increases further. Also the peaked component above 10 Hz becomes more pronounced at the highest energies. Although detailed studies of the energy dependence over the 0.5-10 keV energy of the island state PDS are sparse, the complex energy dependence we have found resembles that what has been found for the hard state PDS in several black-hole systems \citep[e.g.,][]{2009MNRAS.397..666W,2013ApJ...766...89K,2015MNRAS.454.2360D}. The hard state of black-hole systems resembles in many aspect the island state of atoll sources, so it is not surprising that also the energy dependence in the 0.5-10 keV energy range shows similarities. 

\subsection{\exo}\label{subsection:exo}

Similarly to \citet[][]{2017arXiv170307389M} we found that the PDS of {\exo} (green curve in left panel of Figure \ref{fig:allPDS}) is relatively flat over the full frequency range \citep[in the $\nu P_\nu$ representation, which is different than what was used by][]{2017arXiv170307389M}, although we do find a small decrease in strength at the lowest and highest frequencies. The decrease at high frequencies was not explicitly mentioned by \citet[][]{2017arXiv170307389M},  although it can be seen from their Figure 6 as well. 

The total integrated noise had an amplitude of 38.3$\pm$0.3\% rms in the 0.5-10 keV energy range and it slightly increased at higher photon energy (Table \ref{tab:energydependency}; Figure \ref{fig:energydependency} panel c). We note that in the energy range 0.5-1 keV the source was barely detected (because of the high column density toward the source) and the noise component in the obtained PDS was not significant (hence we do not show it in panel c of Figure \ref{fig:energydependency}). Therefore, the 0.5-10 keV PDS is dominated by the photons in the range 1-10 keV and, as expected, the integrated noise in the 1-10 keV band (Table \ref{tab:energydependency}) is identical to the one in the 0.5-10 keV band. We found that the noise showed nearly an identical shape in the two highest energy bands (Figure \ref{fig:energydependency} panel c).

\subsection{\msource} \label{subsec:m28ipds}

\begin{figure}
 \begin{center}
\includegraphics[width=\columnwidth]{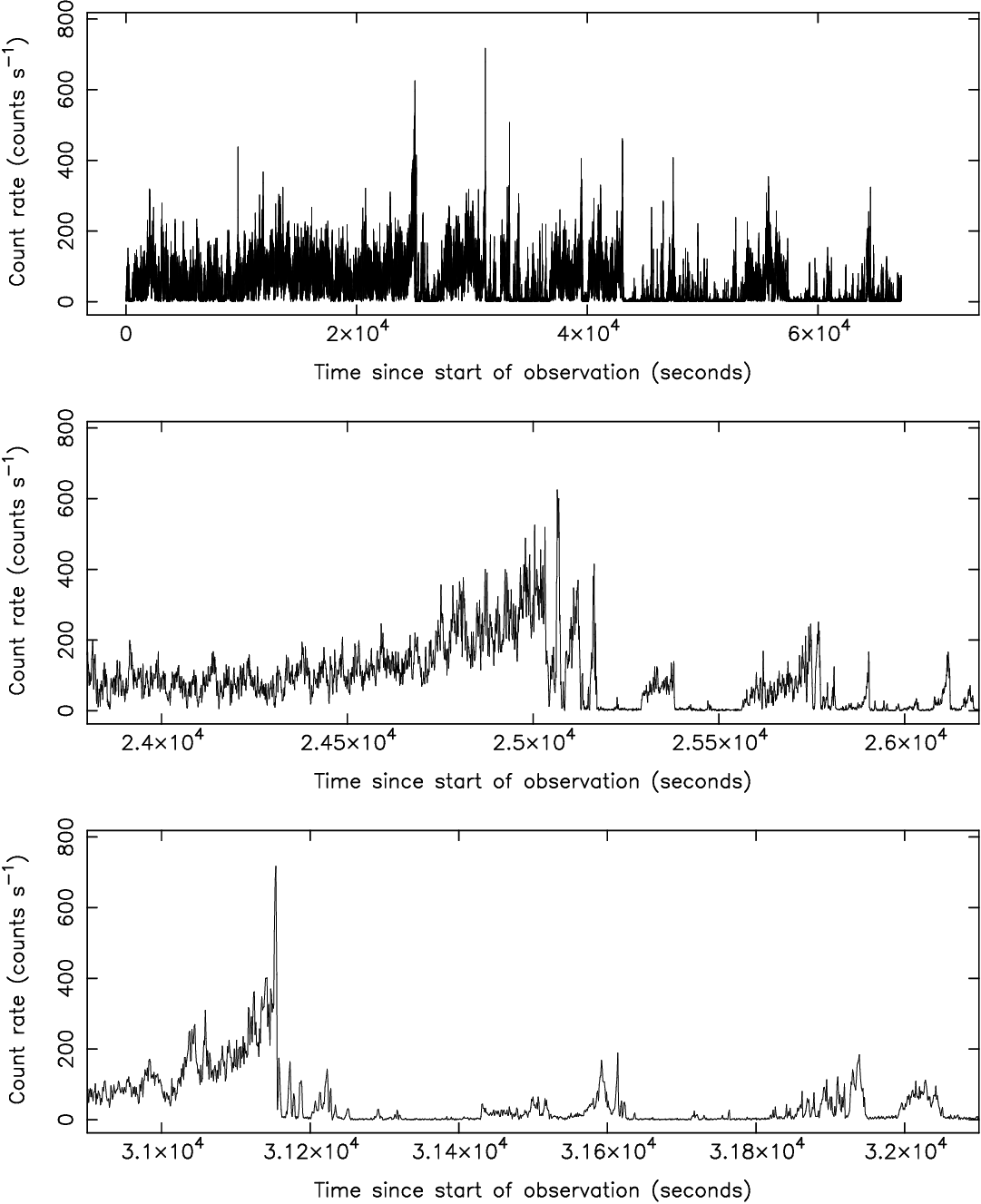}
    \end{center}
 \caption[]{The 1-s light curve of {\msource} (second observation; 0.5-10 keV). Top panel: full light curve of the observation; middle and bottom panel: two zooms of part of the light curve. }
\label{fig:M28ilc}
\end{figure}

The PDS created for the two observations of {\msource} look nearly identical, although the integrated noise was slightly stronger during the first observations compared to the second observation (102.0$\pm$1.9\% versus 91.9$\pm$1.2\% rms amplitude, respectively). Since the second observation was the longest of the two, we only show the PDS of this observation of the source in Figure \ref{fig:allPDS} (dark blue curve in left panel) and Figure \ref{fig:energydependency} (panel d). We note the exceptionally high rms amplitudes during both observations as well as the different shape of the PDS compared to the other two sources so far discussed (Figure \ref{fig:allPDS} left): the noise continues to increase in strength until roughly 0.01 Hz below which it stayed constant in strength \citep[see also Figure 12\footnote{The units of the y-axis (power axis) appear incorrect in Figure 12 of \citet[][]{2014A&A...567A..77F}. They also use the $\nu P_\nu$ representation but their power values for the noise components are $>$1 order of magnitude lower than what we see in Figure \ref{fig:allPDS} (dark blue cruve in left panel).} of][]{2014A&A...567A..77F}. During both observations the shape of the noise in the PDS was approximately the same for all energy bands used in our work (Figure \ref{fig:energydependency}). However, the strength of the noise increased with photon energy (see Table \ref{tab:energydependency}).

The strong variability can clearly also be seen in the light curves of the source presented in Figure \ref{fig:10secondLCs} (panels d and e). To further highlight the extreme variability we show in Figure \ref{fig:M28ilc} the light curve of the second {\it XMM-Newton} observation at a time resolution of 1 second. The top panel of that figure shows the whole observation and the other two panels the parts of the light curve that show the strongest variability. From those two bottom panels it can clearly be seen that the source reached a maximum count rate of $>$600 count s$^{-1}$ (0.5-10 keV) but within $<100$ seconds the count rates reached close to zero. The light curve was also already presented by \citet[][their Figure 1]{2014A&A...567A..77F} but due to their adaptive rebinning, the extremeness of the variability did not show up as clearly as in our Figure \ref{fig:M28ilc}. This variability is of quite a different structure than the variability we see in the other sources (see Figure \ref{fig:10secondLCs}).

\begin{figure}
 \begin{center}
\includegraphics[width=\columnwidth]{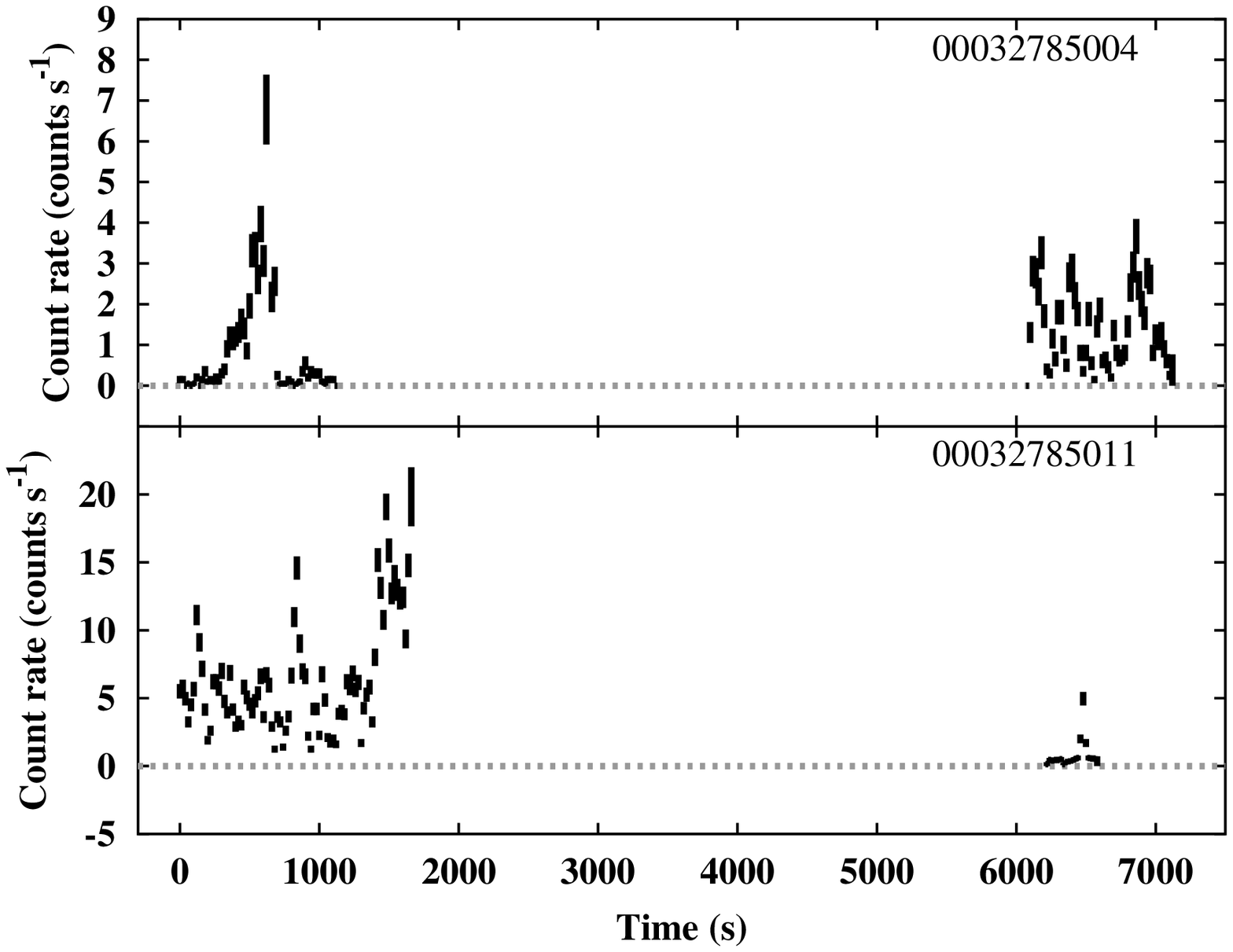}\
    \end{center}
 \caption[]{The {\it Swift}/XRT light curves (20 seconds resolution) for {\msource} as obtained from the observations with ID numbers 00032785004 and 00032785011 showing the strong variability during those two observations. In Figure \ref{fig:m28i_lcs} (left, top sub-panel) it is indicated when these observations occurred during the full outburst of the source. }
\label{fig:m28i_var}
\end{figure}

As can be seen from Figure \ref{fig:m28i_lcs} (left), the {\it XMM-Newton} observations were separated in time by $\sim$9 days (Table \ref{tab:log}). It is likely that the {\it XMM-Newton} observations did not occur at special moments during the outburst but that the extreme variability we observe during those observations was also present throughout the rest of the outburst. To investigate this, we created high time resolution light curves of the {\it Swift}/XRT data presented in Figure \ref{fig:m28i_lcs} (left, top sub-panel). We indeed found that the source was quite variable during those {\it Swift}/XRT observations. We show two clear examples in Figure \ref{fig:m28i_var} (the times when those observations occurred during the outburst are shown in Figure \ref{fig:m28i_lcs}, left).  The type of variability seen during those {\it Swift}/XRT observations resembles that seen during the {\it XMM-Newton} observations. This short term variability could likely also explain the long term variability seen during the whole outburst (Figure \ref{fig:m28i_lcs}, left, top sub-panel).

\subsection{\igr}

The properties (i.e., shape and strength) of the noise components in the PDS of {\jsource} and {\exo} resemble those observed for the strong noise component detected in  the {\it RXTE} PDS for {\igr} obtained during its 2004 outburst \citep[][]{2007ApJ...660..595L}. Similar PDS were observed for this source during its 2015 outburst using {\it XMM-Newton} data by \citet[][]{2017MNRAS.466.3450F}. To be able to compare this source in detail with our other two sources, we show in Figure \ref{fig:allPDS} (light blue curve in the right panel) the {\it XMM-Newton} PDS of {\igr} as well. This PDS clearly shows a very strong noise component. The integrate noise has an rms amplitude of 57.1$\pm$0.3\% (0.5-10 keV; Table \ref{tab:energydependency}). This is significantly larger than what we have  observed for {\jsource} and {\exo}. The shape of this noise component is similarly broad as to what we have observed for the other two sources, although its exact shape differs significantly. The PDS of {\igr} is more complex than observed for the other two sources. It has two broad peaked components at $\sim$0.01 Hz and $\sim$0.3 Hz. On top of this noise component a strong $\sim$8 mHz QPO is present.

The strength of the integrated noise does not have a clear correlation with photon energy (Table \ref{tab:energydependency}), although this might be blurred by the very strong energy dependence of the 8 mHz QPO whose strength decreases very steeply with photon energy. This can be clearly seen in Figure \ref{fig:energydependency}, panel e, from which also can be noted that the noise at frequencies $>$0.01 Hz in general increases in strength at all frequencies with increasing photon energy. The shape of the noise component $>$0.01 Hz does not change significantly with energy. We refer to \citet[][]{2017MNRAS.466.3450F} for a detailed study of {\it XMM-Newton} PDS of {\igr} and the energy dependence of the different components in its PDS. 

\subsection{\ngcsource}

Very recently \citet[][]{2017ApJ...837...61B} suggested that the {\it XMM-Newton} PDS he obtained for {\ngcsource} during its 2016/2017 outburst resembled the PDS of {\igr}. Both sources have very strong noise with similar very low characteristics frequencies. To compare both sources in detail with each other and with the other sources in our sample, we show the two PDS (corresponding to the two {\it XMM-Newton} observations available) of {\ngcsource} in Figure \ref{fig:allPDS} as well (pink and brown curves in the right panel). Broadly speaking, the PDS of {\ngcsource} show a strong noise component over a broad frequency range that peaks at $\sim$0.1 Hz and $\sim$10 Hz. The noise seen during the second observation has slightly lower characteristic frequencies than the noise component observed during the first observation. However, as shown by \citet[][]{2017ApJ...837...61B} the noise components during both observations are more complex (including some QPOs) and we refer to that paper for a detailed study of the PDS of {\ngcsource}.

The 0.5-10 keV integrated noise has an rms amplitude of 26.2$\pm$0.7\% during the first observation and 30.9$\pm$0.7\% during the second observation. For both observations, this noise strongly increases with photon energy (see Table \ref{tab:energydependency}. This can also be seen in panel f of Figure \ref{fig:energydependency} (only the second, longest, observation is shown), from which also can be seen that the overall shape does not change significantly with photon energy (although we note that for frequencies $>$1 Hz the statistical quality of the PDS is low). The difference in strength and the characteristic frequencies of the noise components between the two {\it XMM-Newton} observations might be related to the fact that the second observation was taken at slightly lower luminosities than the first observation (see Figure \ref{fig:j0911_lcs}, third panel).

When comparing the PDS of {\ngcsource} with the other sources, it is clear that indeed the PDS resembles those observed for the other sources (not taking into account the extreme behaviour of {\msource} and the second observation of {\jsource}) with respect to the strength of the noise and its characteristic frequencies. However, it is also obvious that the noise in {\ngcsource} is the weakest in our sample and has the highest characteristics frequencies. In addition {\ngcsource} resembles more {\jsource} and {\exo} than {\igr}, i.e., when comparing the strength of the noise component.

\section{Discussion}\label{sec:discuss}

\citet[][]{parikh2017} identified very hard spectra at relatively  high X-ray luminosities (between $10^{36}$ and $10^{37}$ erg s$^{-1}$ in the range 0.5-10 keV; corresponding to roughly 1\%-10\% of the Eddington luminosity for a neutron star) in three neutron-star LMXBs (\jsource, \exo, \msource). They suggested that the spectra indicate a distinct spectral state in these sources compared to the ones previously identified in neutron-star LMXBs when they have similar luminosities. Although commonly states in such systems are indeed classified based on their X-ray spectra alone (although regularly also including the luminosity of the sources), originally states in neutron-star LMXBs were defined based on their correlated spectral-timing behaviour \citep[see][]{1989A&A...225...79H}. Therefore, in this paper we have investigated the rapid X-ray variability of the above mentioned sources using {\it XMM-Newton} observations that were performed during the period when the sources where in their proposed very hard state.

\begin{figure}
 \begin{center}
\includegraphics[width=\columnwidth]{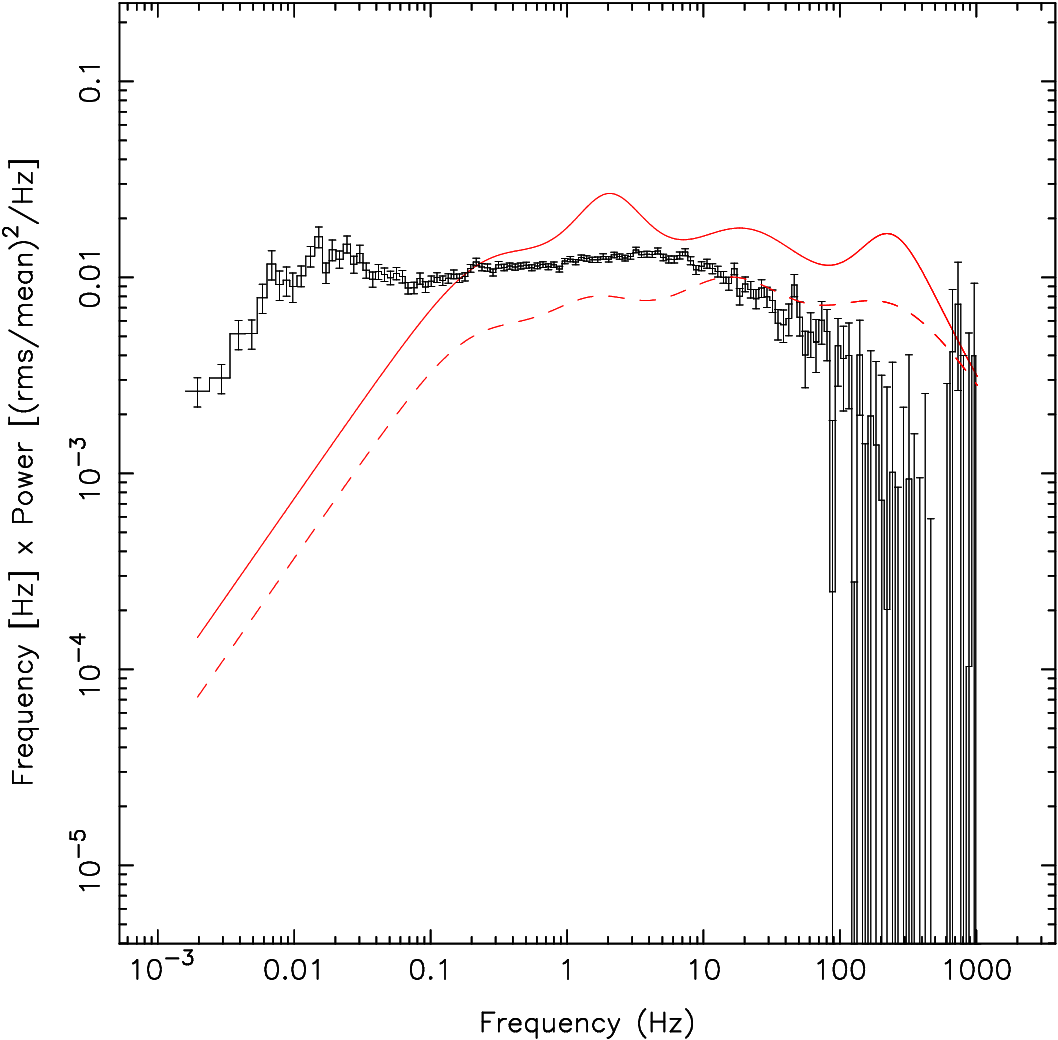}\
    \end{center}
 \caption[]{The PDS of {\jsource} obtained during its very hard state (i.e., obtained during its first {\it XMM-Newton} observation; note the PDS is now shown up to a frequency of 1024 Hz compared to 128 Hz in Figure \ref{fig:allPDS}). The solid red line indicates the PDS observed during the extreme islands state of 4U 0614+09 \citep[][panel 1 in their Figure 2]{2002ApJ...568..912V}; the dashed red line indicates the extreme island state PDS of 4U 1608-52 \citep[][panel A in their Figure 7]{2003ApJ...596.1155V}.}
\label{fig:1804EIS}
\end{figure}

We created PDS of our sources and our results presented in Section \ref{sec:timing} demonstrate that very likely we have indeed identified a new spectral-timing state in {\jsource} and {\exo} (Section \ref{subsec:distinctstate}). The behaviour of {\msource} is very complex and it is unclear how its fits in in any state classification of neutron-star LMXBs (Section \ref{subsec:m28i}). We also compared our results with that of the AMXPs {\igr} and {\ngcsource} for which previously similar X-ray variability phenomena have been observed. We suggest that despite that these two sources did not exhibited as hard spectra as seen for the other sources, that they might also exhibited variability characteristics related to the very hard state behaviour we see in the other sources (Section \ref{subsec:igr}). 

\subsection{A distinct spectral-timing state in {\jsource} and {\exo}}\label{subsec:distinctstate}

{\jsource} and {\exo} show a lot of similarities when they are in their very hard state. When comparing the light curves and photon index curves displayed in the left and right panels in Figure \ref{fig:1804_lcs} it is clear that the outburst profiles of both sources are very similar. Both stay in a quasi-stable state for 1 to 2 months at a luminosity between $10^{36}$ and $10^{37}$ erg s$^{-1}$ before they exhibited the brightest part of their outbursts (where the outburst luminosity peaked around a factor 10 higher; i.e., the transition into the soft, banana branch state). During this quasi-stable state the spectra for both sources were very hard (as measured in the 0.5-10 keV range) with photon indices between 1 and 1.5 \cite[see also][]{parikh2017}. 

In addition, for both sources we found that they exhibited very strong variability over a large range of frequencies (see Figure \ref{fig:allPDS}, left panel). The rms amplitudes over $9.765625 \times 10^{-4}$-128 Hz were $\sim$34\%-38\% (in the range 0.5-10 keV). These properties resemble what is seen in atoll sources when they are in their extreme island state during which also very strong noise is observed \citep[up to 30\%-40\% rms amplitude; e.g.,][]{1997ApJ...485L..37M,2002ApJ...568..912V}. However, typically the characteristic frequencies of the noise during the extreme island state are higher (by a factor $>$10) than we observed for our two sources. To compare more directly the PDS during the very hard state of our sources with the PDS seen during the extreme island state in atoll sources, we show in Figure \ref{fig:1804EIS} the very hard state PDS of {\jsource} again together with the shape of the extreme island state PDS observed for 4U 0614+09 \citep[][]{2002ApJ...568..912V} and 4U 1608-52 \citep[][]{2003ApJ...596.1155V}. It can be seen that during the extreme island state the noise shape and strength is broadly similar to what we observe for the very hard state PDS of {\jsource}, but shifted to higher frequencies. This could suggest that the very hard state behaviour is just an extension of the extreme island state to harder spectra and lower characteristics frequencies. However, the luminosities observed during the very hard state are very similar to those seen during both the island and the extreme island states in atoll sources. Therefore, it remains unclear how these states are exactly related to the very hard state we have idenfitied.

For {\jsource} we obtained another PDS using a second {\it XMM-Newton} observation performed later on in the outburst (see Fig.~\ref{fig:1804_lcs} left). This observation was performed only 1 to 2 days before the source transited to its soft (banana branch) state \citep[which was around April 3, 2015; see][]{2016MNRAS.461.4049D}. During this observation the luminosity and the photon index were only slightly larger but the PDS looked remarkably different (see Fig. \ref{fig:allPDS}, left panel). This PDS looks remarkably similar to what is observed for the island state of atolls sources and we suggest that {\jsource} was indeed in this state during its second {\it XMM-Newton} observation. This suggest that only a minor increase of luminosity (and thus assumed increase in accretion rate; factor of $\sim$2; see Section \ref{sec:bursts}) caused the source to transit from the very hard state to the hard/island state. However, only a minor change (softening) in spectral shape (between 0.5-10 keV) was observed\footnote{The photon index observed during the {\it Swift}/XRT observations performed closest in time to the {\it XMM-Newton} observations increased from 1.10$\pm$0.03 around the time of the first {\it XMM-Newton} observation to 1.31$\pm$0.02 around the time of the second one (see Figure \ref{fig:1804_lcs} right).}. Therefore, despite that the very hard state was first identified based only on spectral information \citep[][]{parikh2017}, such spectral studies alone will most often not be enough to distinguish between the very hard state and the hard state in a specific source. The rapid X-ray variability properties need to be studied as well before definitive conclusions can be made. 

Despite the many similarities between {\jsource} and {\exo} in their very hard states, there are some subtle differences. As can be seen from Figure \ref{fig:1804_lcs} (left), {\jsource} was rather stable in X-ray luminosity and spectral hardness during its very hard state. In contrast {\exo} was less stable in its X-ray luminosity and, in particular, in its measured spectral hardness (Figure \ref{fig:1804_lcs}, right). However, during the 2000 outburst of {\exo}, the source showed very similar behaviour (i.e., similar outburst profile and rapid X-ray variability properties) but in the very hard state during this outburst a lot of X-ray dips were observed \citep[see][for a detailed study of this outburst using  {\it RXTE} data]{altamirano2017}. These dips strongly affected the X-ray spectra of the source and it is feasible that similar dipping behaviour was present during the very hard state of the 2015 outburst of the source. During its {\it XMM-Newton} observation, we did observe strong variability but not exactly like the dipping behaviour seen during its 2000 outburst \citep[see Figure \ref{fig:10secondLCs}; see also Figure 5 of ][for an additional zoom of the light curve of {\exo}]{2017arXiv170307389M}. 

\begin{figure}
 \begin{center}
\includegraphics[width=\columnwidth]{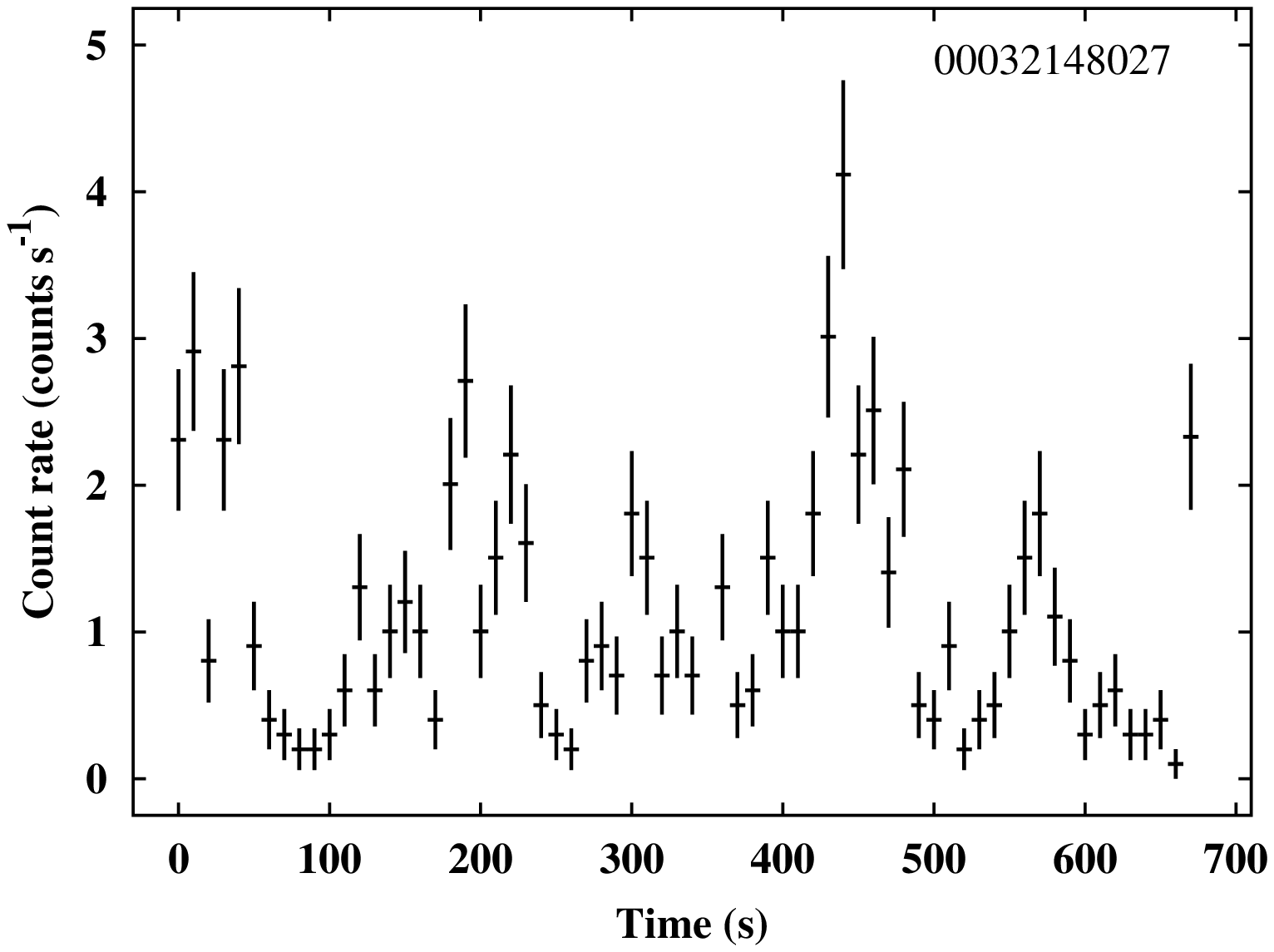}\
    \end{center}
 \caption[]{The {\it Swift}/XRT light curve (10 seconds resolution) for {\exo} as obtained from the observation with ID number 00032148027 showing the strong variability during this observation. In Figure \ref{fig:1804_lcs} (right, third sub-panel) it is indicated when this observation occurred during the full outburst of the source.}
\label{fig:exo1745_var}
\end{figure}

However, as shown by \citet[][]{altamirano2017} not all light curves during the very hard state in this source exhibited strong dipping so the {\it XMM-Newton} observation might have occurred during such an instance. During other episodes of its 2015 outburst, the source might still have showed dipping behaviour. Therefore, we also created higher time resolution light curves of our {\it Swift}/XRT observations taken during the very hard state and indeed during some observations very strong variability was observed. In Figure \ref{fig:exo1745_var} we show the best example which coincided with the lowest luminosity point during the very hard state (as indicated in the third sub-panel of Fig. \ref{fig:1804_lcs} right).  Clearly, very strong variability is present (i.e., stronger than we observe during the {\it XMM-Newton} observation) which indeed could be due to dipping behaviour. Even more so since Figure \ref{fig:exo1745_var} looks very similar to some of the dipping light curves presented by \citet[][]{altamirano2017}. Since we combined all data per {\it Swift}/XRT observation in one luminosity and spectral point for Figure \ref{fig:1804_lcs} (right), we think that it is rather plausible that dipping occurred during (some of) those observations as well and this would obviously effect the final determined luminosity and photon index. Therefore, one has to be cautious when using the reported luminosities and, in particular, the photon indices at face value.

Although we did not see the characteristics of the dipping behaviour during our {\it XMM-Newton} observation of {\exo} it cannot be excluded that some dips were present as well but that the other strong variability we detected blurred our ability to distinguish the dips. However, the presence of dips could possibly be the reason why at the lowest frequencies in the PDS, {\exo} behaves slightly different from {\jsource}: {\exo} remains more flat at the lowest frequencies compared to {\jsource}. Since the dips occur on relatively long time scales, this could be the reason for the extra noise in the PDS. We would like to note that similar PDS were reported by \citet[][]{altamirano2017}, both during periods of strong dipping and when the dipping has ceased. The PDS obtained during the latter phase strongly demonstrate that the strong variability seen in {\exo} is not solely due to the strong dipping behaviour but that the accretion flow intrinsically is very variable and the dips only contribute extra variability at the lowest frequencies.

The dips seen in {\exo} are likely due to some region in the disc that regularly obscures the inner part of the accretion disc and the neutron star. This would suggest that {\exo} has a relatively high inclination. However, very recently, \citet[][]{2017arXiv170307389M} estimated only a rather low inclination of $\sim37^\circ$ for the source using broad band spectral modelling and it is currently unclear how these rather different inclination estimates can be reconciled with each other \citep[see for a more detailed discussion][]{altamirano2017}.  However, if indeed the dips are related to the inclination of the system, the fact that we do not see such dips in {\jsource} likely indicates that the source is at a lower inclination than {\exo}. However, we cannot exclude that the mechanism behind the dips in {\exo} was simply absent in {\jsource}, which might be a distinct possibility since we do not understand their origin.

We would also like to note that \citet[][]{altamirano2017} found that the spectra during the very hard state of this source during its 2000 outburst where indeed significantly harder than what was observed for the source during the other states. The outburst was studied using {\it RXTE} so there was no coverage below $\sim$3 keV, making direct spectral comparisons with our results non-trivial. \citet[][]{altamirano2017} used CDs and HIDs to study the spectral variations and clearly the very hard state had significantly larger hard colours than the other states. In particular, in the CD and HID the very hard state was dislocated from the regular atoll track traced out by the source during the remainder of the outburst. In addition, the very hard state had higher luminosities than the island state which was later observed in this source during the decay phase of its 2000 outburst (i.e., the very hard state was not observed during the decay). This again demonstrates that the luminosity behaviour of atoll sources is complex and at similar luminosities different energy spectra and PDS can be observed. As final point we note that the fact that {\exo} exhibited its very hard state during its 2000 and 2015 outburst demonstrate that this state is a recurrent feature of this source and not a very rare event. Therefore, the chances are high that during its next outburst this very hard state can be observed again.  However, we note that for both outbursts the very hard state was only observed during the initial phase of the outbursts.

From the comparison between {\jsource} and {\exo} we conclude that both sources exhibited spectral-timing states that lasted $\sim$2 and $\sim$1 month, respectively, and they are indeed distinct from the island and extreme island states often seen in atoll sources. Although in {\exo} it is a recurrent phenomena and has now been observed twice, the question is why this very hard state was not previously identified in other neutron-star LMXBs. The observed luminosities are relatively high ($10^{36-37}$ erg s$^{-1}$) and relatively easily accessible in many previous studies of such systems. This strongly suggest that such very hard states are not very common in neutron-star LMXBs in general and that a (possibly rare) source specific property is needed for a particular system to exhibit this state. 
 
\subsection{{\msource}}\label{subsec:m28i}

Despite the fact that the 0.5-10 keV X-ray spectra of {\msource} are as hard as those observed for {\jsource} and {\exo} (see Figures \ref{fig:1804_lcs}, \ref{fig:m28i_lcs}, and \ref{fig:gammavsLx}), the rapid X-ray variability of {\msource} is quite different and even more extreme than observed for the other two sources. In Figure \ref{fig:M28ilc} we show an example of the variability observed for this source. The count rate sometimes goes up to $>$600 counts s$^{-1}$ and then down to nearly zero within a time span of several tens of seconds. Such rapid flaring causes the very high rms amplitude of the noise in the PDS and likely is also responsible for its shape (see Figure \ref{fig:allPDS}, left panel). This flaring behaviour was already noticed by \citet[][see their Figure 1]{2014A&A...567A..77F}. They suggested a “hiccup” form of accretion in which the system alternates between a state in which the accretion of matter onto the neutron star is halted by the neutron-star magnetic field (and possibly causing matter outflow from the system) and a state in which the matter can reach the neutron-star surface. Since this system is an AMXP this mechanism might indeed be present, although it is unclear why only this AMXP would show this behaviour and none of the other AMXPs.

{\jsource} and {\exo} do not show obvious pulsations in the  {\it XMM-Newton} data \citep[see, e.g.,  the detailed pulse search performed for {\exo} by][]{2017arXiv170307389M} and no pulsations could be seen in the {\it RXTE} data of {\exo} either \citep[][]{altamirano2017}\footnote{The upper limits on the signal amplitude on any pulsations  in {\exo} were between a few percent and 15 percent, depending on the method used and the assumptions made on the spin and orbital parameters of the source \citep[][]{2017arXiv170307389M}. A similar pulse search for {\jsource} would likely result in comparable upper limits but such a detailed study is beyond the scope of our paper. }. Although pulsations could be still present in those two sources, it is striking to note that when a particular AMXP exhibits millisecond X-ray pulsations, those oscillations always show up very readily in its PDS (i.e., so far {\it all} AMXPs were discovered when making simple PDS without the need for sophisticated pulse search methods). In contrast, the high frequency PDS ($>100$ Hz) of {\jsource} and {\exo} did not show any sign at all for pulsations. This strongly suggest that both sources do not pulsate when they are accreting (at least they are not typical AMXPs when in outburst) indicating that the neutron-star magnetic fields in those systems do not play a very important role in the accretion processes. 

If the hiccup accretion seen in {\msource} is indeed due to the strong role of the magnetic field in this system, it might be possible that this form of accretion would also produce very hard spectra. Similar very hard spectra were observed for neutron-star LMXBs that harbour neutron stars with strong magnetic fields \citep[$10^{11-12}$ G; e.g., for the sources 4U 1626-67 and GRO J1744-28; see, e.g.,][]{1997A&A...324L...9O,2012A&A...546A..40C,2015MNRAS.452.2490D,2015ApJ...804...43Y}. However, the magnetic fields in these systems are significantly stronger than the $10^{8-9}$ G field estimated for the neutron star in {\msource}. This estimate is very similar to what is inferred for other AMXPs that do not exhibit very hard spectra (see Sections~\ref{subsec:igr} and~\ref{section:AMXPs}). Moreover, the 11 Hz pulsar in the globular cluster Terzan 5 (IGR J17480-2446) has an intermediate strong magnetic field  with an estimated strength of $10^{9-10}$ G but its X-ray spectra are not very hard either \citep[e.g.,][]{2011MNRAS.418..490C,2012MNRAS.423.1178P}. Therefore, there appears not to be a clear relationship between the magnetic field strength and the hardness of the spectra. However, the hiccup accretion mechanism in {\msource} so far seems to be a very unique feature in this source. It might be possible that the magnetic field in this system affects the accretion flow in a different way than what is observed in the other AMXPs (as well as in IGR J17480-2446) and potentially this difference could also cause the very hard spectra in {\msource}.  If indeed the magnetic field in {\msource} plays a role, then its very hard spectra  are created differently than in the other two, non-pulsating sources and two mechanisms could be active in (weak-field) neutron-star LMXBs that could generate such very hard spectra. The behaviour of {\msource} might then indicate another state in neutron-star LMXBs as speculated upon by \citet[][]{2014A&A...567A..77F}. With the current available data we cannot conclusively determine whether or not the very hard state in {\msource} is related to the very hard states observed in {\jsource} and {\exo}.

\subsection{{\igr} and {\ngcsource}}\label{subsec:igr}

The strength of the noise in the PDS of {\jsource} and {\exo} and how it is distributed over a broad frequency range resembles that seen for the AMXP {\igr} \citep[][]{2007ApJ...660..595L} during its 2004 outburst. This behaviour was already pointed out by \citet[][]{2007ApJ...660..595L} as being anomalous compared to the PDS typically seen for atoll sources (i.e., in the island and extreme island states). Those authors suggested that the source was in an ``exceptional island state'' at a luminosity where one would typically expect to observe island or extreme island state PDS in atoll sources. Since the source exhibited in 2015 another outburst which was also observed with {\it XMM-Newton} \citep[][]{2017MNRAS.466.3450F}, we were able to include this source in our study as well. As already pointed out by \citet[][]{2017MNRAS.466.3450F}, the PDS of {\igr} during its 2015 outburst resembles closely the ones seen by \citet[][]{2007ApJ...660..595L}\footnote{\citet[][]{2017MNRAS.466.3450F} reported the discovery of a strong QPO at $\sim$8 mHz which was not seen by \citet[][]{2007ApJ...660..595L}. This is likely due to the fact that the QPO strength decreases strongly with energy. Above the lower-energy threshold of the {\it RXTE}/PCA ($\sim$3 keV), the QPO may simply have been too weak to be detectable \citep[see also Figure \ref{fig:energydependency} and the discussion in][]{2017MNRAS.466.3450F}. We note that very recently a tentative $\sim$6 mHz QPO was reported in the {\it RXTE} data by \citet[][]{2017arXiv170706445B} and this QPO, if confirmed, could indeed be related to the $\sim$8 mHz QPO seen in the {\it XMM-Newton} data.}, demonstrating that the behaviour during both outbursts was very similar. 

Indeed, we find a noise component in the PDS that is spread out over a similarly large frequency range as seen in {\jsource} and {\exo}, although the noise in the PDS for {\igr} is considerably stronger than what we oberved for the other two sources. In addition, the shape of the PDS differs in the details. However, this was already the case when comparing the PDS of {\jsource} and {\exo} so it is unclear how important the detailed shapes of the PDS (including its energy dependence) are when comparing different sources. To compare the hardness of the spectra for {\igr} with our other two sources, we created Figure \ref{fig:gammavsLx}, in which we show the photon index (measured for the 0.5-10 keV spectra) versus the 0.5-10 keV X-ray luminosity \citep[after][]{2015MNRAS.454.1371W,parikh2017}. Despite the fact that {\igr} is not as hard as {\jsource} and {\exo} in the luminosity range $10^{36-37}$ erg s$^{-1}$, the source is still harder than the other neutron-star LMXBs in this figure (interestingly, this can also be inferred from Figure 4 of \citet[][]{2007MNRAS.378...13G} in which {\igr} was one of the hardest sources in their sample). Therefore, it is plausible that the state {\igr} was in is related to the very hard state of {\jsource} and {\exo}. 

Recently \citet[][]{2017ApJ...837...61B} suggested that the PDS obtained (using {\it XMM-Newton} data) during the 2016 outburst of another AMXP, {\ngcsource}, resembled that of {\igr}. Therefore, we also included this source in our analysis and we confirm the noise properties reported for this source by \citet[][]{2017ApJ...837...61B}. Although it is clear from Figure \ref{fig:allPDS} that the source does indeed exhibit strong noise in its PDS, it is not as strong as we observe for {\jsource} and {\exo}, let alone what we observed for {\igr}. In addition, the characteristic frequencies observed for {\ngcsource} are significantly higher than for the other three sources and are somewhere in between the frequencies observed for those sources and those observed in the classical extreme island state in other atoll sources (see Figure \ref{fig:1804EIS}). 

The PDS of {\ngcsource} showed an evolution with luminosity: when the luminosity decreased slightly, the noise increased in strength while its characteristic frequencies decreased (Figures \ref{fig:1804EIS} and  \ref{fig:j0911_lcs}; Table \ref{tab:energydependency}). This noise behaviour is similar to the noise seen for atoll sources in their island and extreme island state and could signify that the different noise components might be connected. Several other observable phenomena might strengthen this conclusion: (a) the noise in {\ngcsource} and the island state noise observed for {\jsource} during its second {\it XMM-Newton} observation have the strongest energy dependency (Table \ref{tab:energydependency}), and (b) when plotting the photon index of {\ngcsource} versus the X-ray luminosity (Figure \ref{fig:gammavsLx}) the source is fully consistent with the other neutron-star LMXBs which are presumed to be in the [extreme] island state at these luminosities. However, we caution about drawing too strong inferences from such comparisons because we have only a few sources to compare with each other and still the noise of {\ngcsource} has rather low characteristic frequencies compared to other extreme island state sources. The state {\ngcsource} was in during its observation might be more related to the very hard state we observed for our other sources than to the extreme island state observed in the normal atoll sources.

\subsection{Connection with AMXPs} \label{section:AMXPs}

When looking at our source sample, it is remarkable that three of our five targets are AMXPs. We have already have argued in Section \ref{subsec:m28i} that {\jsource} and {\exo} are unlikely to be AMXPs as well and therefore the unusual state of our sources is likely not solely related to the sources being an AMXPs (i.e., having clear evidence for the presence of a dynamically important neutron-star magnetic field). Moreover, {\msource} is so extreme and unusual that it was probably in yet another state, so far only observed in this source. Despite this, many rapid X-ray variability studies have been performed for many accreting neutron stars, and the AMXPs appear over-represented in our sample. 

Assuming that indeed AMXPs are more likely to display this unusual accretion state, then the magnetic field might indeed play an important role in this. For example, the disc might be truncated at relatively large radii by the magnetic field and therefore the variability might have low characteristic frequencies (see the discussions in \citet[][]{2007ApJ...660..595L} and \citet[][]{2017ApJ...837...61B}). We note, however, that this ``truncated disc'' hypothesis cannot explain the very hard state in {\jsource} since during this state it was found that the inner disk was close to the neutron star as reported by \citet[][an inner disk radius of $\leq22.2$ km was reported by these authors]{2016ApJ...824...37L}. In addition, it remains then unclear why out of $\sim$16 AMXPs, only {\igr} and {\ngcsource} (excluding for now the extreme behaviour of {\msource}) exhibit this unusual state. \citet[][]{2017ApJ...837...61B} compared in detail those two sources with what is seen for other AMXPs \citep[for typical PDS of other AMXPs see, e.g.,][]{2005ApJ...619..455V} and he also could not find an observational property (such as spin frequency, orbital period, outburst properties) that would distinguish those two sources from the general AMXPs population. Clearly, more systems (both AMXPs and non-pulsating neutron-star LMXBs) need to be found and studied in detail to find correlation of the occurrence (or absence) of this state with other source properties. In addition, more detailed studies (at all wavelengths) need to be performed to investigate the physical nature of this very hard state. Since {\ngcsource} is still active at the time of submission our paper, this source would be an excellent target right now to perform additional observations during this state of the source to further elucidate the nature of this state.

\section*{Acknowledgements}

RW and AP acknowledge support from a NWO Top Grant, Module 1, awarded to RW. ND acknowledges support from an NWO Vidi grant. DA acknowledges support from the Royal Society. This research has made use of NASA's  Astrophysics Data System, of the {\it MAXI} data provided by RIKEN, JAXA and the {\it MAXI} team, and of the {\it Swift}/BAT transient monitor results provided by the {\it Swift}/BAT team.

\bibliographystyle{mn2e}

\begin{thebibliography}{}



\bibitem[Altamirano et al.(2005)]{2005ApJ...633..358A} Altamirano, D., van der Klis, M., M{\'e}ndez, M., et al.\ 2005, \apj, 633, 358 

\bibitem[Altamirano et al.(2008)]{2008ApJ...685..436A} Altamirano, D., van der Klis, M., M{\'e}ndez, M., et al.\ 2008, \apj, 685, 436-450 


\bibitem[Altamirano et al.(2015)]{2015ATel.7240....1A} Altamirano, D., Krimm, H.~A., Patruno, A., et al.\ 2015, The Astronomer's Telegram, 7240  

\bibitem[Altamirano et al.(2017)]{altamirano2017} Altamirano, D., et al. 2017, MNRAS, in prepration

\bibitem[Barret \& Olive(2002)]{2002ApJ...576..391B} Barret, D., \& Olive, J.-F.\ 2002, \apj, 576, 391 


\bibitem[Belloni et al.(1997)]{1997A&A...322..857B} Belloni, T., van der Klis, M., Lewin, W.~H.~G., et al.\ 1997, \aap, 322, 857 

\bibitem[Bildsten(1998)]{1998ASIC..515..419B} Bildsten, L.\ 1998, NATO Advanced Science Institutes (ASI) Series C, 515, 419 


\bibitem[Bult(2017)]{2017ApJ...837...61B} Bult, P.\ 2017, \apj, 837, 61 

\bibitem[Bult et al.(2017)]{2017arXiv170706445B} Bult, P., van Doesburgh, M., \& van der Klis, M.\ 2017, ApJ, in press (arXiv:1707.06445) 


\bibitem[Camero-Arranz et al.(2012)]{2012A&A...546A..40C} Camero-Arranz, A., Pottschmidt, K., Finger, M.~H., et al.\ 2012, \aap, 546, A40 

\bibitem[Chakraborty et al.(2011)]{2011MNRAS.418..490C} Chakraborty, M., Bhattacharyya, S., \& Mukherjee, A.\ 2011, \mnras, 418, 490 



\bibitem[De Falco et al.(2017a)]{2017A&A...599A..88D} De Falco, V., Kuiper, L., Bozzo, E., et al.\ 2017a, \aap, 599, A88 


\bibitem[De Falco et al.(2017b)]{2017arXiv170404181D} De Falco, V., Kuiper, L., Bozzo, E., et al.\ 2017b, \aap, 603, A16


\bibitem[Degenaar et al.(2016)]{2016MNRAS.461.4049D} Degenaar, N., Altamirano, D., Parker, M., et al.\ 2016, \mnras, 461, 4049 

\bibitem[De Marco et al.(2015)]{2015MNRAS.454.2360D} De Marco, B., Ponti, G., Mu{\~n}oz-Darias, T., \& Nandra, K.\ 2015, \mnras, 454, 2360 

\bibitem[Doroshenko et al.(2015)]{2015MNRAS.452.2490D} Doroshenko, R., Santangelo, A., Doroshenko, V., Suleimanov, V., \& Piraino, S.\ 2015, \mnras, 452, 2490 


\bibitem[Eckert et al.(2013)]{2013ATel.4925....1E} Eckert, D., Del Santo, M., Bazzano, A., et al.\ 2013, The Astronomer's Telegram, 4925  

\bibitem[Ferrigno et al.(2014)]{2014A&A...567A..77F} Ferrigno, C., Bozzo, E., Papitto, A., et al.\ 2014, \aap, 567, A77 

\bibitem[Ferrigno et al.(2017)]{2017MNRAS.466.3450F} Ferrigno, C., Bozzo, E., Sanna, A., et al.\ 2017, \mnras, 466, 3450 

\bibitem[Fridriksson et al.(2015)]{2015ApJ...809...52F} Fridriksson, J.~K., Homan, J., \& Remillard, R.~A.\ 2015, \apj, 809, 52 

\bibitem[Fujimoto et al.(1981)]{1981ApJ...247..267F} Fujimoto, M.~Y., Hanawa, T., \& Miyaji, S.\ 1981, \apj, 247, 267 

\bibitem[Galloway et al.(2005)]{2005ApJ...622L..45G} Galloway, D.~K., Markwardt, C.~B., Morgan, E.~H., Chakrabarty, D., \& Strohmayer, T.~E.\ 2005, \apjl, 622, L45 

\bibitem[Galloway et al.(2008)]{2008ApJS..179..360G} Galloway, D.~K., Muno, M.~P., Hartman, J.~M., Psaltis, D., \& Chakrabarty, D.\ 2008, \apjs, 179, 360-422 

\bibitem[Gierli{\'n}ski \& Done(2002)]{2002MNRAS.331L..47G} Gierli{\'n}ski, M., \& Done, C.\ 2002, \mnras, 331, L47 

\bibitem[Gladstone et al.(2007)]{2007MNRAS.378...13G} Gladstone, J., Done, C., \& Gierli{\'n}ski, M.\ 2007, \mnras, 378, 13 

\bibitem[Harris(1996)]{1996AJ....112.1487H} Harris, W.~E.\ 1996, \aj, 112, 1487 

\bibitem[Hasinger \& van der Klis(1989)]{1989A&A...225...79H} Hasinger, G., \& van der Klis, M.\ 1989, \aap, 225, 79 

\bibitem[Homan et al.(2004)]{2004A&A...418..255H} Homan, J., Wijnands, R., Rupen, M.~P., et al.\ 2004, \aap, 418, 255 

\bibitem[Homan et al.(2010)]{2010ApJ...719..201H} Homan, J., van der Klis, M., Fridriksson, J.~K., et al.\ 2010, \apj, 719, 201 

\bibitem[Kalamkar et al.(2013)]{2013ApJ...766...89K} Kalamkar, M., van der Klis, M., Uttley, P., Altamirano, D., \& Wijnands, R.\ 2013, \apj, 766, 89 

\bibitem[Klein-Wolt \& van der Klis(2008)]{2008ApJ...675.1407K} Klein-Wolt, M., \& van der Klis, M.\ 2008, \apj, 675, 1407-1423 

\bibitem[Krimm et al.(2013)]{2013ApJS..209...14K} Krimm, H.~A., Holland, S.~T., Corbet, R.~H.~D., et al.\ 2013, \apjs, 209, 14 

\bibitem[Krimm et al.(2015)]{2015ATel.6997....1K} Krimm, H.~A., Barthelmy, S.~D., Baumgartner, W., et al.\ 2015, The Astronomer's Telegram, 6997  

\bibitem[Kuster et al.(1999)]{1999SPIE.3765..673K} Kuster, M., Benlloch, S., Kendziorra, E., \& Briel, U.~G.\ 1999, Proc. SPIE, 3765, 673 

\bibitem[Kuster et al.(2002)]{2002astro.ph..3207K} Kuster, M., Kendziorra, E., Benlloch, S., et al.\ 2002, arXiv:astro-ph/0203207 

\bibitem[Linares et al.(2007)]{2007ApJ...660..595L} Linares, M., van der Klis, M., \& Wijnands, R.\ 2007, \apj, 660, 595 

\bibitem[Linares et al.(2014)]{2014MNRAS.438..251L} Linares, M., Bahramian, A., Heinke, C., et al.\ 2014, \mnras, 438, 251 

\bibitem[Ludlam et al.(2016)]{2016ApJ...824...37L} Ludlam, R.~M., Miller, J.~M., Cackett, E.~M., et al.\ 2016, \apj, 824, 37 

\bibitem[Matranga et al.(2017)]{2017arXiv170307389M} Matranga, M., Papitto, A., Di Salvo, T., et al.\ 2017, \aap, 603, A39 

\bibitem[Matsuoka et al.(2009)]{2009PASJ...61..999M} Matsuoka, M., Kawasaki, K., Ueno, S., et al.\ 2009, \pasj, 61, 999 

\bibitem[M{\'e}ndez et al.(1997)]{1997ApJ...485L..37M} M{\'e}ndez, M., van der Klis, M., van Paradijs, J., et al.\ 1997, \apjl, 485, L37 

\bibitem[Muno et al.(2002)]{2002ApJ...568L..35M} Muno, M.~P., Remillard, R.~A., \& Chakrabarty, D.\ 2002, \apjl, 568, L35 

\bibitem[Ng et al.(2010)]{2010A&A...522A..96N} Ng, C., D{\'{\i}}az Trigo, M., Cadolle Bel, M., \& Migliari, S.\ 2010, \aap, 522, A96 

\bibitem[Oosterbroek et al.(1995)]{1995A&A...297..141O} Oosterbroek, T., van der Klis, M., Kuulkers, E., van Paradijs, J., \& Lewin, W.~H.~G.\ 1995, \aap, 297, 141 



\bibitem[Owens et al.(1997)]{1997A&A...324L...9O} Owens, A., Oosterbroek, T., \& Parmar, A.~N.\ 1997, \aap, 324, L9 

\bibitem[Papitto et al.(2012)]{2012MNRAS.423.1178P} Papitto, A., Di Salvo, T., Burderi, L., et al.\ 2012, \mnras, 423, 1178 


\bibitem[Papitto et al.(2013)]{2013Natur.501..517P} Papitto, A., Ferrigno, C., Bozzo, E., et al.\ 2013, \nat, 501, 517 

\bibitem[Parikh et al.(2017)]{parikh2017} Parikh, A.~S., Wijnands, R., Degenaar, N., et al.\ 2017, \mnras, 468, 3979

\bibitem[Remillard et al.(2006)]{2006ATel..696....1R} Remillard, R.~A., Lin, D., ASM Team at MIT, \& NASA/GSFC 2006, The Astronomer's Telegram, 696  

\bibitem[Sanna et al.(2017a)]{2017A&A...598A..34S} Sanna, A., Papitto, A., Burderi, L., et al.\ 2017a, \aap, 598, A34 

\bibitem[Sanna et al.(2017b)]{2017MNRAS.466.2910S} Sanna, A., Pintore, F., Bozzo, E., et al.\ 2017b, \mnras, 466, 2910 

\bibitem[Schnerr et al.(2003)]{2003A&A...406..221S} Schnerr, R.~S., Reerink, T., van der Klis, M., et al.\ 2003, \aap, 406, 221 

\bibitem[Str{\"u}der et al.(2001)]{2001A&A...365L..18S} Str{\"u}der, L., Briel, U., Dennerl, K., et al.\ 2001, \aap, 365, L18 

\bibitem[Tetarenko et al.(2016)]{2016MNRAS.460..345T} Tetarenko, A.~J., Bahramian, A., Sivakoff, G.~R., et al.\ 2016, \mnras, 460, 345 

\bibitem[Tomsick et al.(2004)]{2004ApJ...601..439T} Tomsick, J.~A., Kalemci, E., \& Kaaret, P.\ 2004, \apj, 601, 439 

\bibitem[van der Klis(2000)]{2000ARA&A..38..717V} van der Klis, M.\ 2000, \araa, 38, 717 

\bibitem[van der Klis(2006)]{2006csxs.book...39V} van der Klis, M.\ 2006, Compact stellar X-ray sources, 39, 39 

\bibitem[van Straaten et al.(2002)]{2002ApJ...568..912V} van Straaten, S., van der Klis, M., di Salvo, T., \& Belloni, T.\ 2002, \apj, 568, 912 

\bibitem[van Straaten et al.(2003)]{2003ApJ...596.1155V} van Straaten, S., van der Klis, M., \& M{\'e}ndez, M.\ 2003, \apj, 596, 1155 

\bibitem[van Straaten et al.(2005)]{2005ApJ...619..455V} van Straaten, S., van der Klis, M., \& Wijnands, R.\ 2005, \apj, 619, 455 

\bibitem[Wijnands(2001)]{2001AdSpR..28..469W} Wijnands, R.\ 2001, Advances in Space Research, 28, 469 

\bibitem[Wijnands \& van der Klis(1999a)]{1999ApJ...514..939W} Wijnands, R., \& van der Klis, M.\ 1999a, \apj, 514, 939 

\bibitem[Wijnands \& van der Klis(1999b)]{1999ApJ...522..965W} Wijnands, R., \& van der Klis, M.\ 1999b, \apj, 522, 965 

\bibitem[Wijnands et al.(1999)]{1999ApJ...512L..39W} Wijnands, R., van der Klis, M., \& Rijkhorst, E.-J.\ 1999, \apjl, 512, L39 

\bibitem[Wijnands et al.(2015)]{2015MNRAS.454.1371W} Wijnands, R., Degenaar, N., Armas Padilla, M., et al.\ 2015, \mnras, 454, 1371 

\bibitem[Wilkinson \& Uttley(2009)]{2009MNRAS.397..666W} Wilkinson, T., \& Uttley, P.\ 2009, \mnras, 397, 666 

\bibitem[Younes et al.(2015)]{2015ApJ...804...43Y} Younes, G., Kouveliotou, C., Grefenstette, B.~W., et al.\ 2015, \apj, 804, 43 


\end{thebibliography}

\appendix 

\section{Outburst light curves} \label{sec:outbursts}

\begin{figure*}
 \begin{center}
\includegraphics[width=\columnwidth]{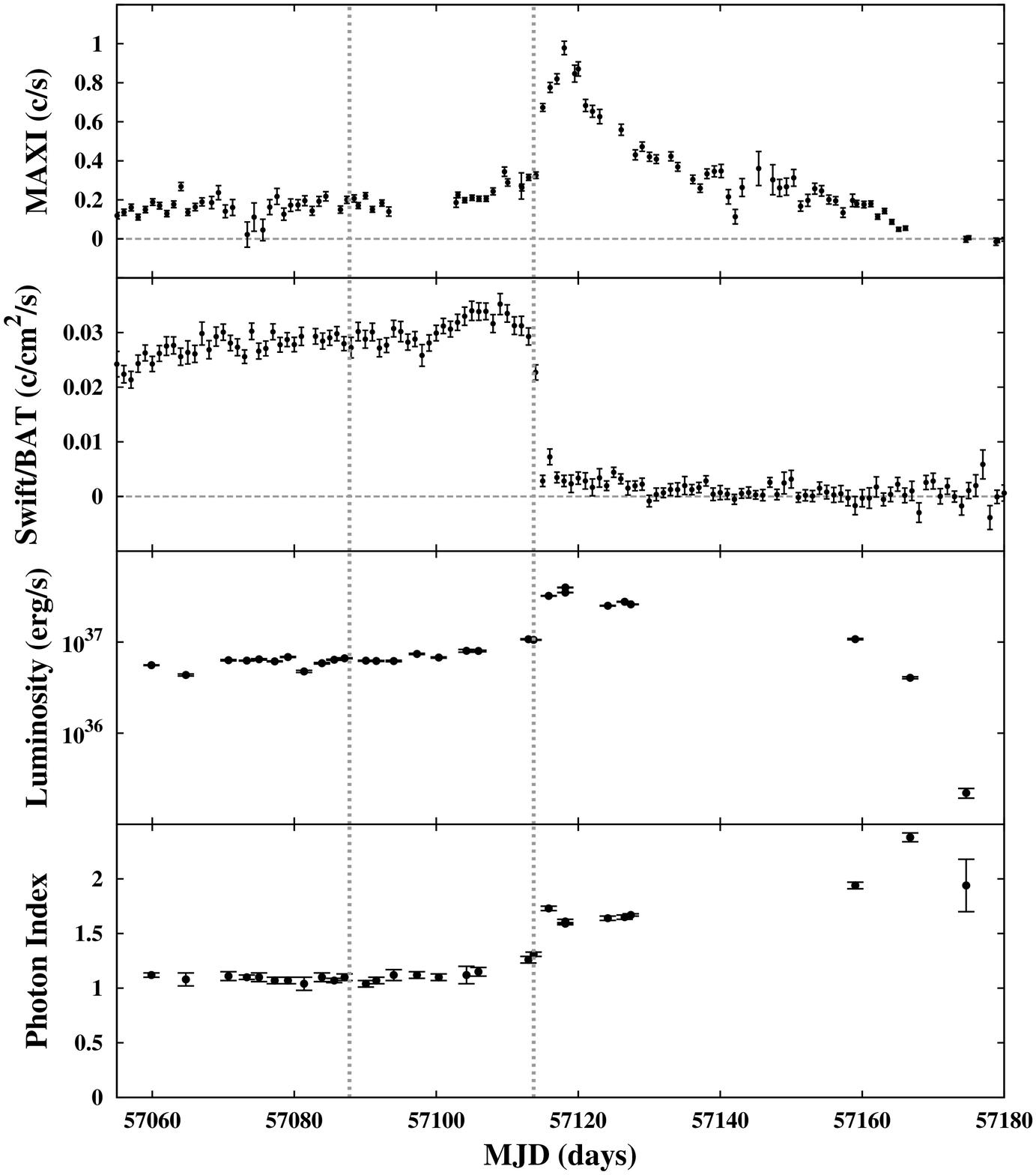}\hspace{0.5cm}
\includegraphics[width=\columnwidth]{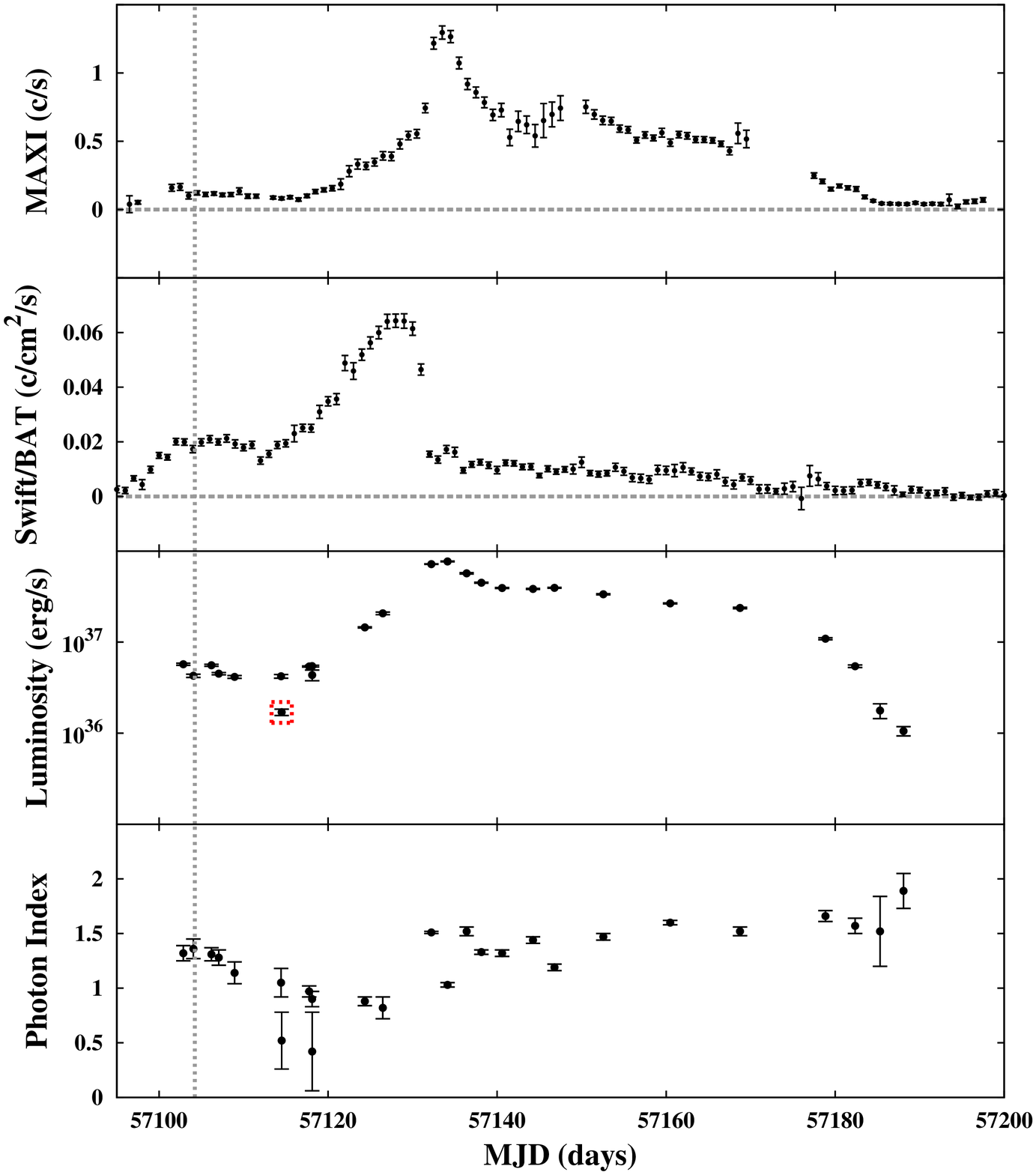}
    \end{center}
 \caption[]{The light curves observed for {\jsource} (left) and {\exo} (right) as obtained during the 2015 outbursts of the sources: {\it MAXI} (top panels; for the energy range 2-10 keV), {\it Swift}/BAT (second panels; 15-50 keV), and {\it Swift}/XRT (third panels; 0.5-10 keV). In the bottom panels we show the evolution of the photon indices versus time during these outbursts as measured using the {\it Swift}/XRT data. The dotted lines indicate the time of the {\it XMM-Newton} observations used in our paper. The {\it MAXI} and BAT data are rebinned to have 1 data point every day. The red dashed square around one of the points in the {\it Swift}/XRT light curve in the third panel for {\exo} indicates which observation was used to create Figure \ref{fig:exo1745_var}.}
\label{fig:1804_lcs}
\end{figure*}

\begin{figure*}
 \begin{center}
\includegraphics[width=\columnwidth]{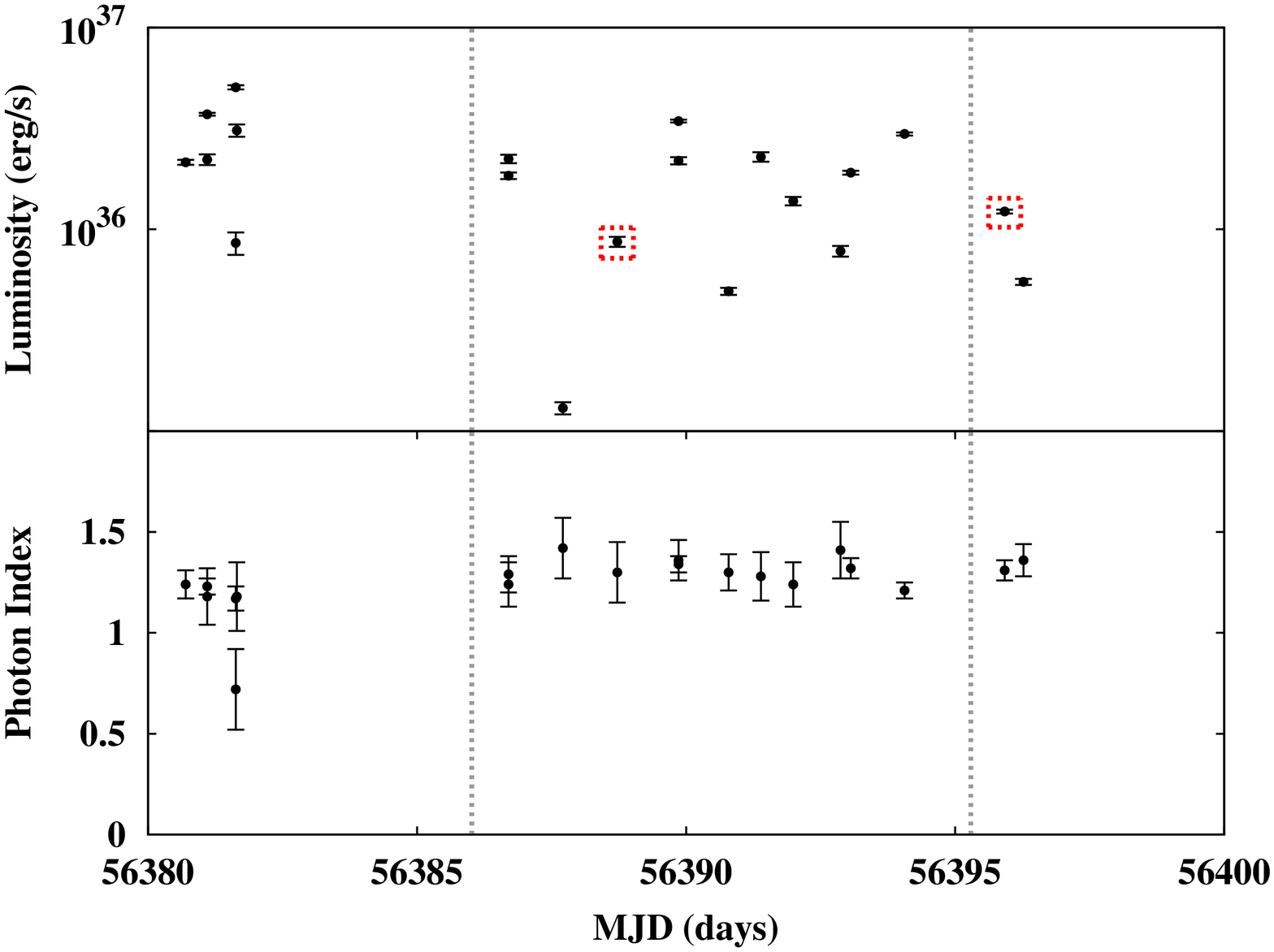}\hspace{0.5cm}
\includegraphics[width=\columnwidth]{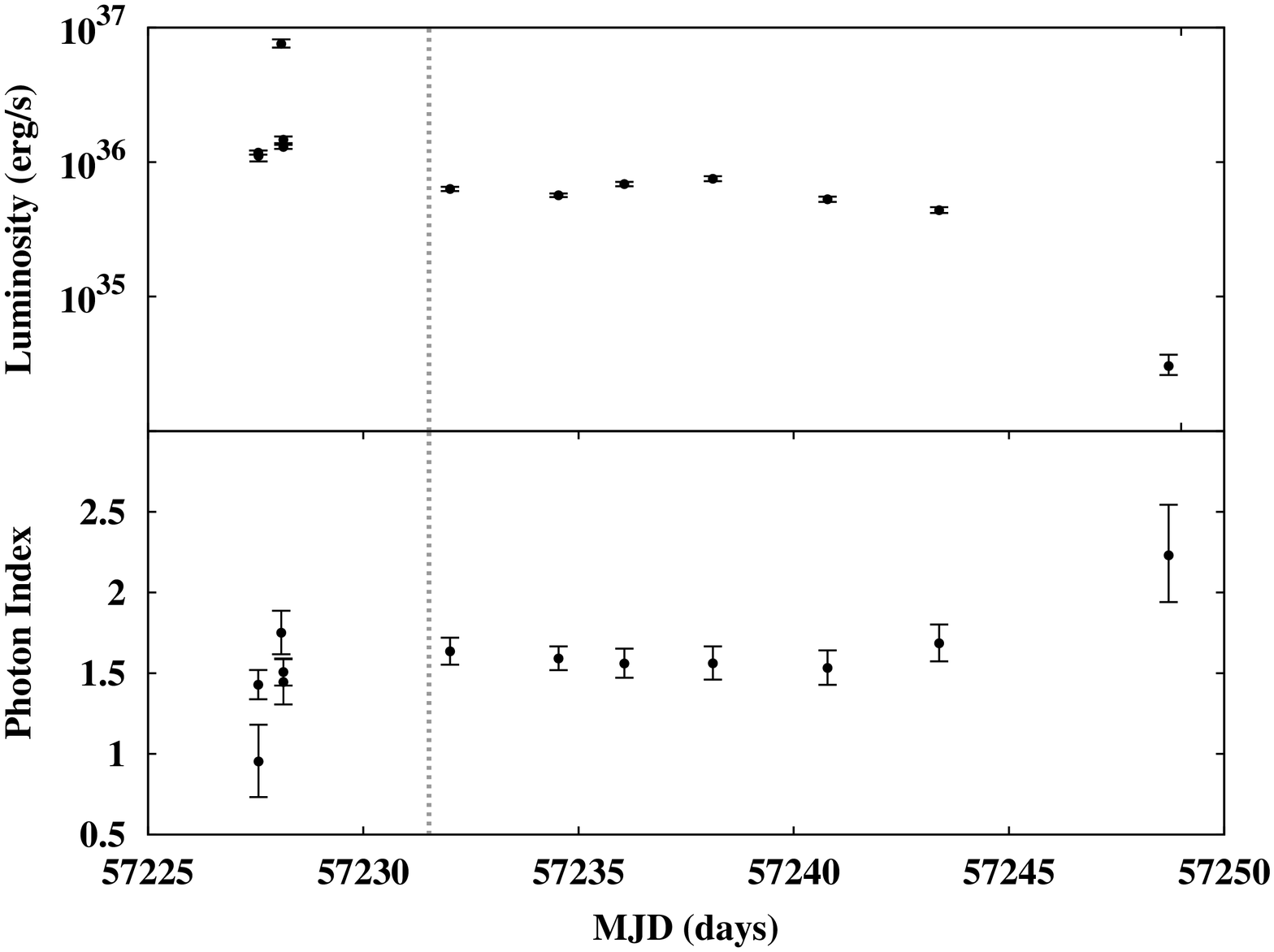}
    \end{center}
 \caption[]{The light curves (top panels; 0.5-10 keV) and photon index curves (bottom panels) obtained for {\msource} (left) and {\igr} (right) using {\it Swift}/XRT observations obtained during the 2013 and 2015 outbursts of the sources, respectively. The dotted lines indicate the time of the {\it XMM-Newton} observations used in our paper. The red dashed square around one of the points in the {\it Swift}/XRT light curve of {\msource} (top panel) indicates which observation we used to create Figure \ref{fig:m28i_var}. We do not show the {\it MAXI} and  {\it Swift}/BAT light curves of these two sources because they have so far not been detected by those instruments, even when the sources were in outburst.}
\label{fig:m28i_lcs}
\end{figure*}

\begin{figure}
 \begin{center}
\includegraphics[width=\columnwidth]{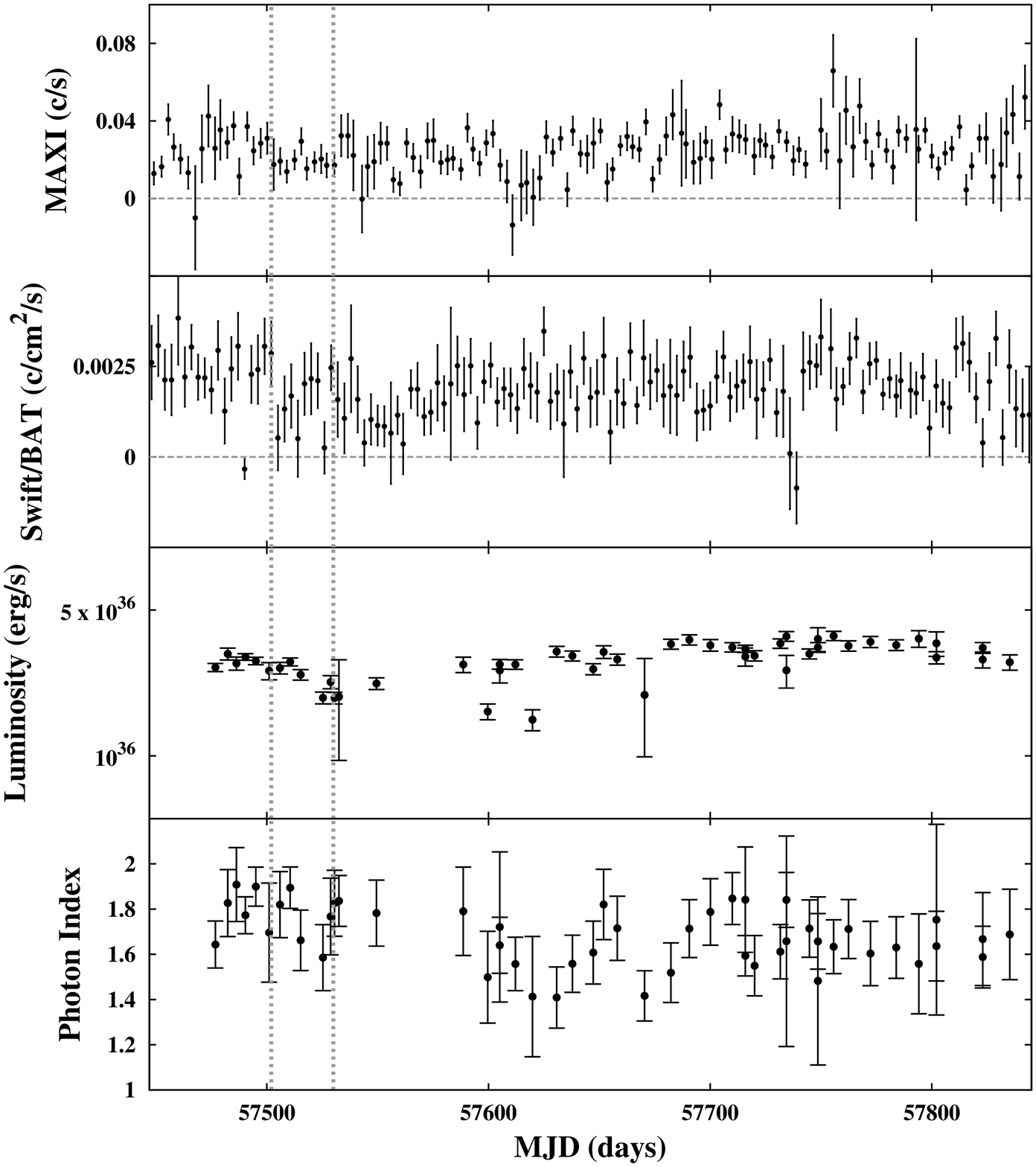}\
    \end{center}
 \caption[]{Similar as to Figure \ref{fig:1804_lcs} but then for {\ngcsource} (during the 2016-2017 outburst of this source). The {\it MAXI} and BAT data are rebinned to have 1 point every 3 days.}
\label{fig:j0911_lcs}
\end{figure}

In Figures \ref{fig:1804_lcs}-\ref{fig:j0911_lcs} we show for all our sources the outburst light curves as well as the photon index curves versus time. The {\it MAXI} data\footnote{http://maxi.riken.jp/top/slist.html} \citep[][]{2009PASJ...61..999M} are for the energy range 2-10 keV and the {\it Swift}/BAT data\footnote{https://swift.gsfc.nasa.gov/results/transients/} \citep[][]{2013ApJS..209...14K} are for 15-50 keV. For {\msource} and {\igr} we do not show the {\it MAXI} and {\it Swift}/BAT light curves because the sources were not detected by those instruments. The times of the {\it XMM-Newton} observations used in our paper are indicated by dotted lines.

\begin{figure}
 \begin{center}
\includegraphics[width=\columnwidth]{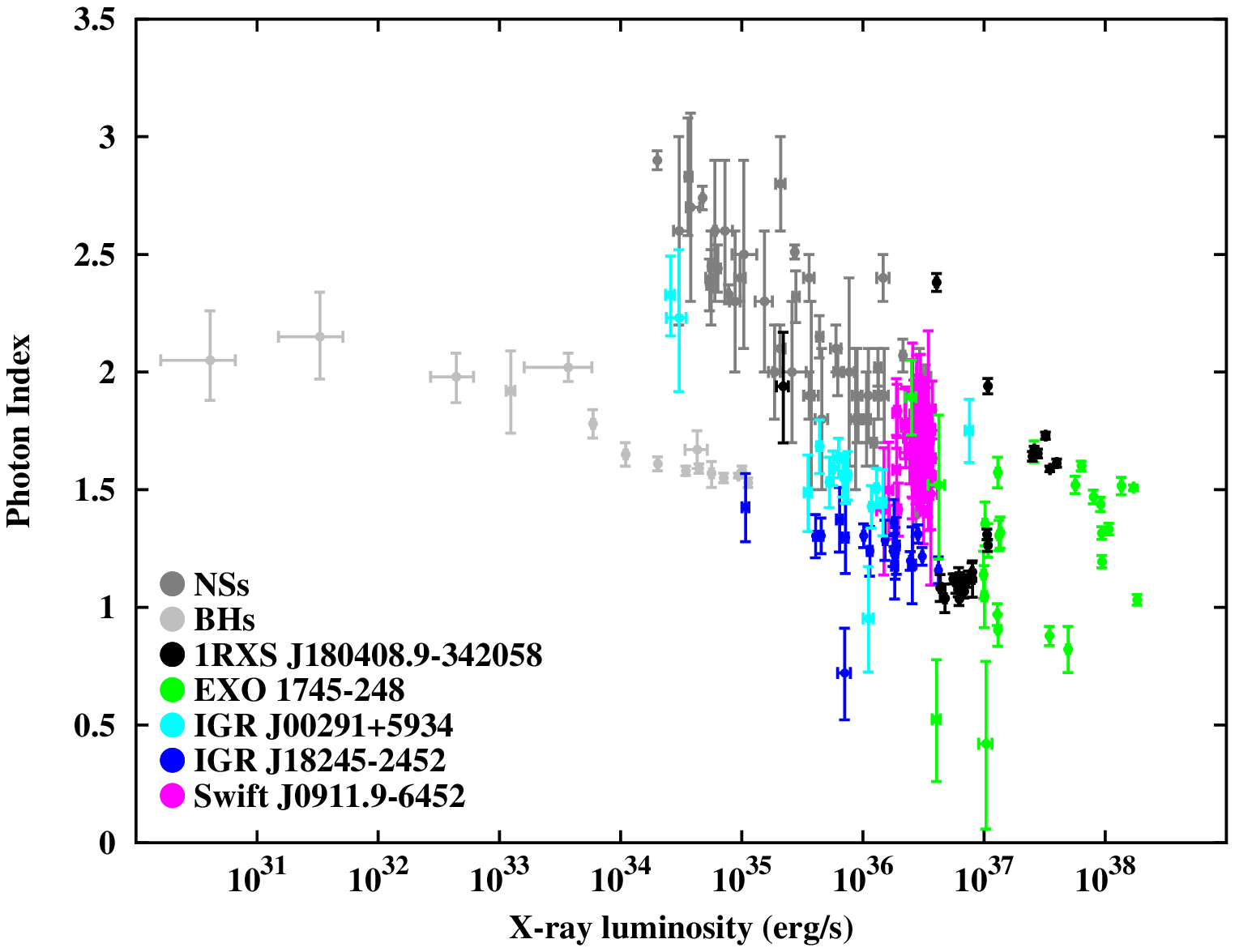}
    \end{center}
 \caption[]{The photon index versus the 0.5-10 keV X-ray luminosity for our five sources. For {\jsource}, {\exo}, and {\msource} we show the same data as presented in \citet[][their Figure 1]{parikh2017}. In addition we have now included  {\igr} and {\ngcsource} for the first time. The grey points (dark grey: the neutron star systems; light grey: the black hole systems) were obtained from \citet[][from their Figure 1]{2015MNRAS.454.1371W}. The colour scheme is the same as in Figure \ref{fig:allPDS} and is indicated in the left bottom corner of the figure.}
\label{fig:gammavsLx}
\end{figure}

The {\it Swift}/XRT results were obtained from \citet[][]{parikh2017} except for {\igr} and {\ngcsource}. For those two sources, we obtained their {\it Swift}/XRT data from the HEASARC data archive\footnote{https://heasarc.gsfc.nasa.gov/} and we processed those data in the same way as was done for the other sources \citep[see details in][]{parikh2017}. To calculate the luminosities we used the same distances used by \citet[][]{parikh2017} for {\jsource} (5.8 kpc), {\exo} (5.5 kpc), and {\msource} (5.5 kpc), and we used a distance of 4 kpc and 9.6 kpc for {\igr} \citep[][]{2005ApJ...622L..45G,2017A&A...599A..88D}  and {\ngcsource} \citep[][2010 update; the source is located in the globular cluster NGC 2808]{1996AJ....112.1487H}, respectively. In Figure \ref{fig:gammavsLx} we also plot the photon index versus X-ray luminosity plot for our sources \citep[see also][]{parikh2017}, including {\igr} and {\ngcsource} for the first time.

\section{Aperiodic variability studies using {\itshape XMM-Newton} EPIC-pn data in timing mode }\label{sec:timingtesting}

When using the {\it XMM-Newton} EPIC-pn camera in timing mode \citep[][]{2001A&A...365L..18S}, one spatial dimension is collapsed  to increase the time resolution of the data (e.g., to avoid pile-up for bright sources or to study their variability properties). The use of this timing mode makes it non-trivial to determine what the optimal (i.e., maximising the SNR of the resulting PDS) source extraction region is and what the background count rate is in this extraction region (needed to renormalise the PDS). 

\begin{figure}
 \begin{center}
\includegraphics[width=\columnwidth]{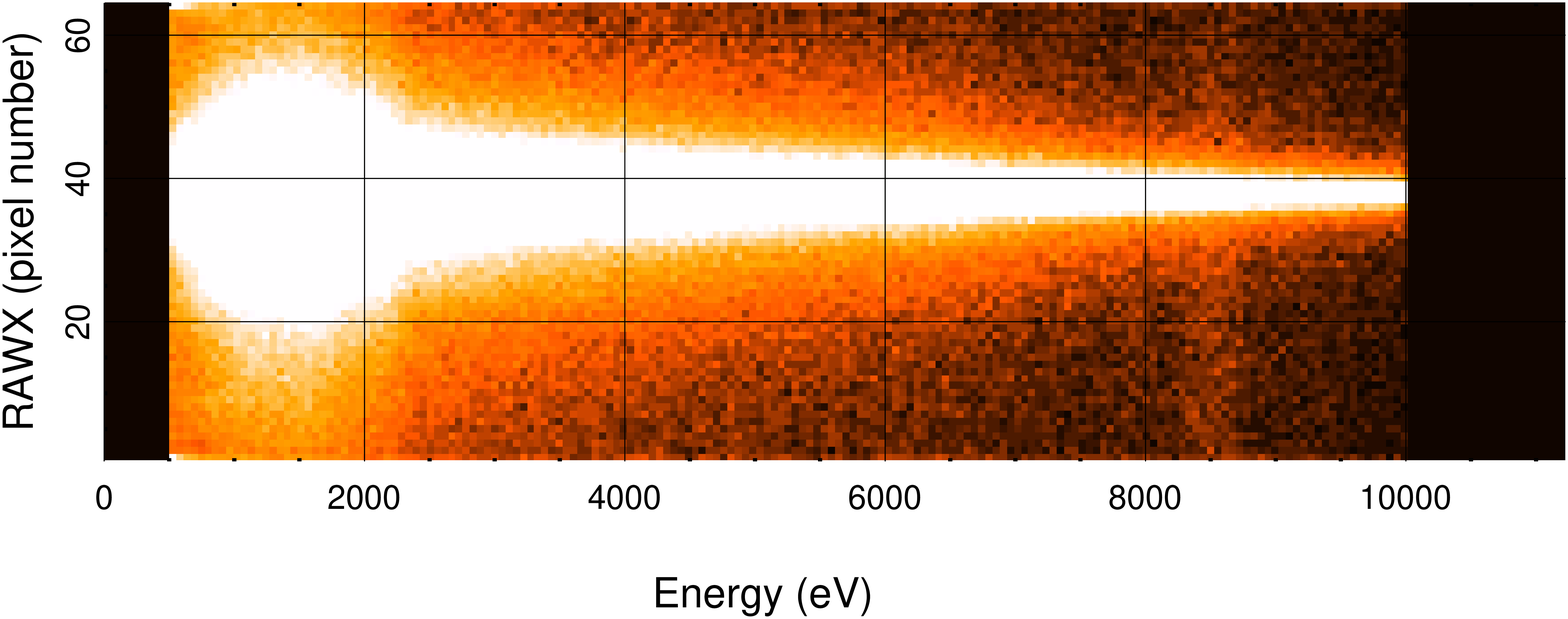}\
    \end{center}
 \caption[]{The RAWX column versus the energy of the detected counts (for the energy range 0.5-10 keV) of the EPIC-pn timing mode data of {\igr}. Clearly, at the lowest energies source counts can be detected all the way to the edges of the CCD. This effects decreases when going to higher energies although for most energies the source contributes significant to the counts over the full RAWX column range. }
\label{fig:RAWXvsPI}
\end{figure}

When the pn is used in timing mode, typically the observed source is (very) bright. Since the extend of point-spread-function of the telescope is such that it is larger than the read-out area of the CCD on which the source falls, the source photons will be spread out over the full read-out area and therefore for bright sources there is no background region that can be used that is free of source photons \citep[e.g., see][see also the {\it XMM-Newton} calibration documents\footnote{E.g., xmm2.esac.esa.int/docs/documents/CAL-TN-0083.pdf}]{2010A&A...522A..96N}.  This effect is clearly visible in Figure \ref{fig:RAWXvsPI} where we show the RAWX column versus the energy of the counts detected for {\igr}. It is clear from this figure that at least up to several keV the source produces a significant amount of counts all the way to the edge of the CCD. Therefore, we cannot extract a source free region to use as background estimate. Since {\igr} is one of the faintest sources in our sample, this effects is even more pronounced for most of the other sources in our sample.

When moving away from the source centre, the number of source counts per RAWX column decreases but the background counts stay approximately constant.  Therefore the SNR in an individual RAWX column decreases when moving to the edge of the camera. It is thus possible that above a certain size of the source extraction region, we could decrease the SNR of the overall PDS if the background count rate starts to dominate over the source count rate for the RAWX columns closer to the edge of the CCD.  This would reduce the accuracy to determine the exact shape of the noise components in the PDS as well as the strength of those components. In addition, the effect of uncertainties in the background count rate estimate increases as well when a larger source extraction region is used.  

Here we study these effects for {\igr} because this source is one of the faintest of our sample so the effects should be largest for this source. Since the observation of this source has one of the longest exposure times and the source exhibited the strongest noise component in its PDS of all our sources, we can obtain excellent constraints on its PDS allowing for small deviations because of the above mentioned effects to become detectable. Also no type-I X-ray bursts were seen during the observation allowing us to use all data for this source. For comparison, we also did the same analysis for the second {\it XMM-Newton} observation of {\jsource} since the source count rate during this observation was the brightest in our sample so the effects should be minimal for these data (we excluded the type-I bursts detected during this observation).

\begin{figure}
 \begin{center}
\includegraphics[width=\columnwidth]{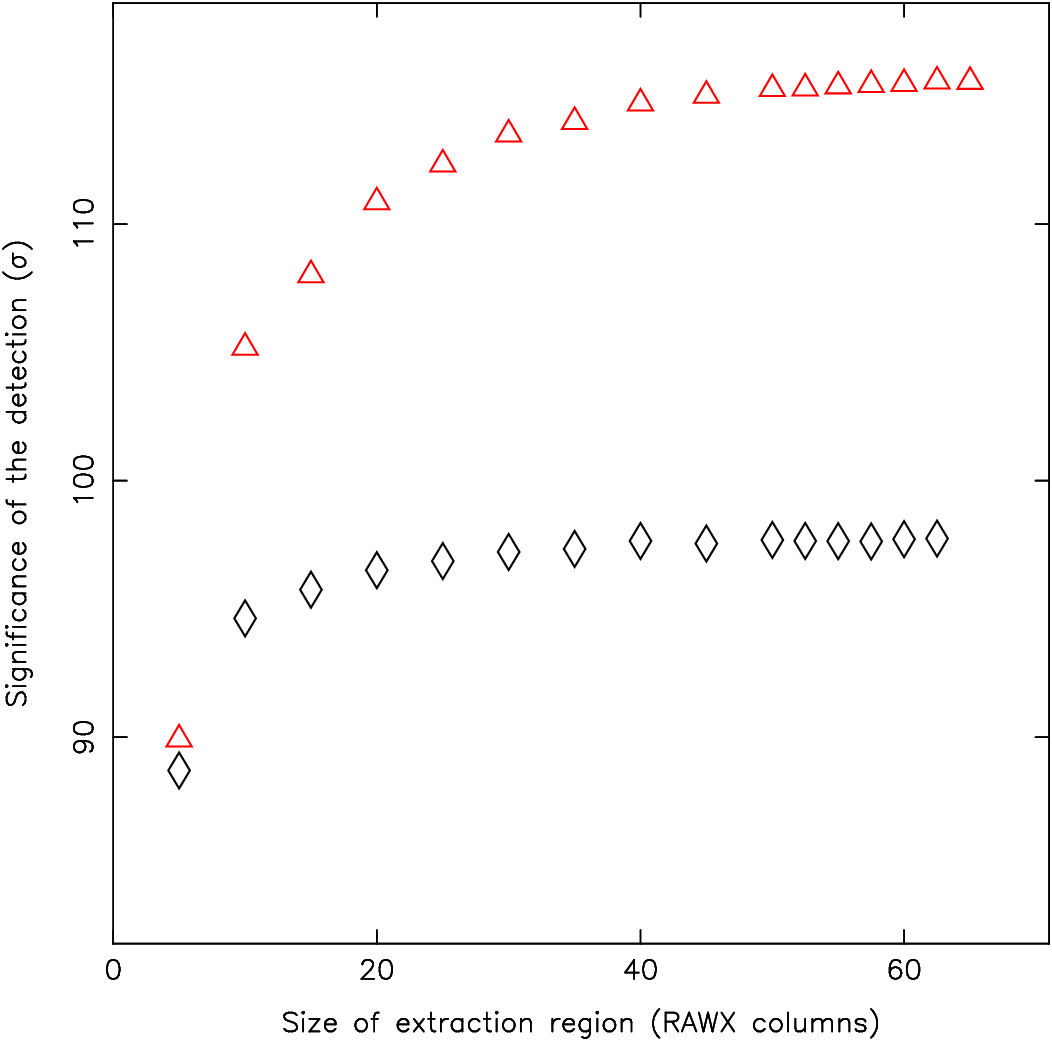}\
    \end{center}
 \caption[]{The significance of the integrated noise (between $9.765625 \times 10^{-4}$ and 128 Hz) for {\igr} (black diamonds) and {\jsource} (red triangles) versus the size of the source extraction region. After a size of 50 RAWX columns the extraction region only increases by half the used width since one side of the CCD has been reached.}
\label{fig:Sigmavssize}
\end{figure}

For both sources, we used source extraction regions (centred on the RAWX column with the highest count rate) that increased in width (in steps of 5 RAWX columns) to study the effect on the PDS, until one boundary of the extraction region reached one side of the CCD (for both sources this occurred when we used an extraction region size of 50 RAWX columns). From that point on we increased the extraction region in the same way but only from one side of the CCD extra data could be extracted (we note that for {\jsource} we had one extraction region more than for {\igr} because the source centres were located at slightly different positions on the CCD). We created 1024 s PDS (for the energy range 0.5-10 keV), subtracted the Poisson level, and renormalised the PDS using a background count rate of zero counts s$^{-1}$ (see Section~\ref{sec:observations} for details about this procedure).  We then integrated the noise in the PDS between $9.765625 \times 10^{-4}$ and 128 Hz ($2^{-10} - 2^{7}$ Hz). We determined the significance of this noise by simply dividing the integrated power by its 1$\sigma$ error. In Figure \ref{fig:Sigmavssize} we plot the obtained significance of the detection versus the size of the source extraction region.

\begin{figure}
 \begin{center}
\includegraphics[width=\columnwidth]{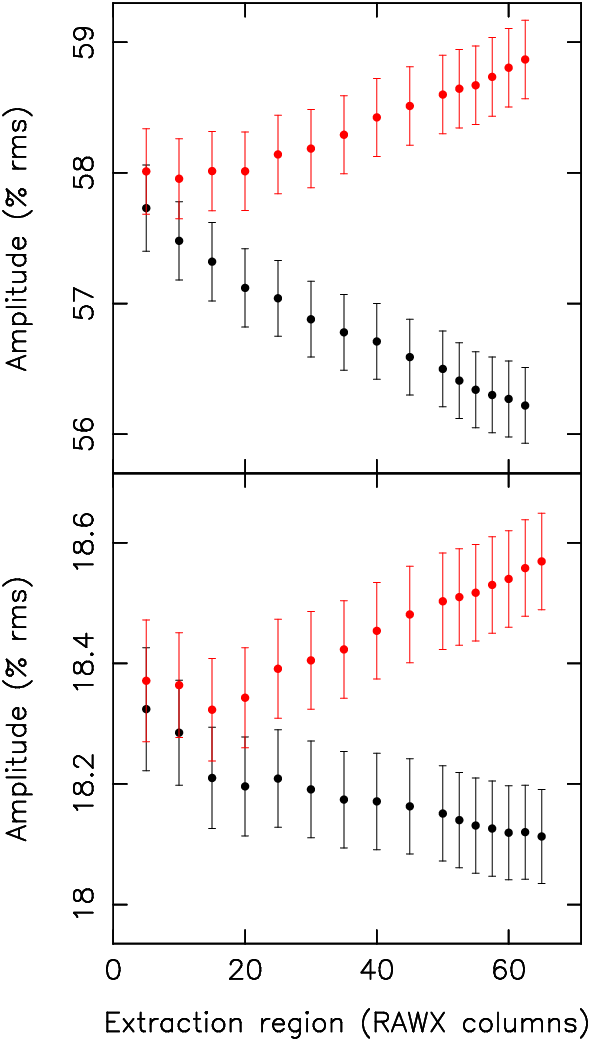}\
    \end{center}
 \caption[]{The strength of the integrated noise ($9.765625 \times 10^{-4}$ - 128 Hz) for {\igr} (top panel) and {\jsource} (bottom panel) versus the size of the extraction region. The red points are calculated when the PDS are corrected for background during the renormalisation using our best estimated for the background count rate (see text); the black points are calculated without any background correction applied to the PDS. After a size of 50 RAWX columns the extraction region only increases by half the size since one side of the CCD has been reached.}
\label{fig:RMSvssize}
\end{figure}

The significance of the noise in the PDS of both sources increases strongly in the beginning with the size of the extraction region and then levels off to a certain maximum significance. Surprisingly, even for our faintest source (\igr) the significance does not decrease for the largest extraction regions indicating that also at the furthest edge of the CCD the source count rate dominate significantly over the background count rate (as can also be seen from Figure \ref{fig:RAWXvsPI}). However, as expected, it is clear that for this source the maximum significance level is reached already at a smaller extraction region than for {\jsource}. In Figure \ref{fig:RMSvssize} (the black points) we plot the rms amplitude of the integrated noise as a function of the extraction region size without correcting for the background count rate. As expected, when the size of the extraction region increases the amplitude of the noise artificially goes down since more background counts are included which do not vary structurally and therefore decreasing the total amount of variability in the PDS. Therefore, the amplitudes shown have to be considered as lower limits to the true values.

To investigate the effects of the uncertainties in the background count rate on the rms amplitude further, we also renormalised the same PDS again but now using our best estimate for the background count rate.  As background region we used a region with a size of 5 RAWX columns centred on a position close to the edge of the CCD (i.e., RAWX = 3; farthest away from the source as possible to minimise the source contribution to the observed count rate in this region). To obtain the background count rate in the different regions we used, the background count rate obtained for this region is then multiplied by the ratio of the size of the source extraction region with the size of the background extraction region. However, since our estimated background count rates are likely an overestimation of the true ones since the source contributes significant to (if not dominates) these obtained count rates. Therefore, the obtained rms amplitude values are upper limits to the true values.

In Figure \ref{fig:RMSvssize} (the red points), we show the resulting rms amplitudes using this estimate background count rate. As expected, the rms values artificially increases with size of the extraction region and are more and more diverging from the true value. This true rms amplitude value must lay between the red and the black points in Figure \ref{fig:RMSvssize}. Since for an extraction region $>$20 RAWX columns the red and the black points start to diverge quickly from each other (for both sources), we decided on an extraction region of 20 RAWX columns for all of our targets in our final analysis to minimise the systematic uncertainties in the calculated rms amplitude values when using a zero background count rate during the renormalisation of the PDS. 

We prefer to obtain (and report) a lower limit (and not an upper limit) on the strength of the observed noise components because that will tell us how strong at least the variability in the data is. Estimating from Figure \ref{fig:RMSvssize}, the systematic uncertainty on the amplitudes is at most $\sim$1\% rms for the weakest sources. By using an extraction region size of 20 RAWX columns, we also obtain already close to the maximum significance reachable for the noise component in the PDS  (see Figure \ref{fig:Sigmavssize}). This is true except for the brightest sources, but then the significance is already very high to start with that we can accurately constrain the shape of the PDS. However, if one is not interested in determining the strength of the component as accurately as possible but is more interested in determining as best as possible the exact shape of the PDS, then one should use a larger extraction region for the brightest sources to increase the SNR of the obtained PDS further.

\subsection{Other effects}

Potentially there could be several other effects that could inhibit us for determining the true strength of the observed noise components in the PDS. Here we discuss two of them briefly. A full investigation of their effects on the calculated rms values is beyond the scope of this paper since they depend on specific properties of the observation used to create the PDS (i.e., the amount of background flaring present as well as the source brightness; both can vary significantly between our observations).

\paragraph*{Effects of background flaring:} Background flaring can cause additional complications in determining the best way to construct the final PDS. Normally one would remove the episodes of strong background flaring but as our targets are relatively bright, even during those episodes of strong flaring the source might still dominate the observed count rate. By removing these episodes of strong flaring we then reduce the SNR of the PDS reducing our ability to determine its shapes and its components most accurately. However, including the periods of elevated background flaring in our PDS will create additional uncertainties on the calculated rms amplitude values since the background count rate contributes additional counts to the observed count rate. Furthermore, if the background flaring is highly variable it could also add to the strength of the observed noise and even alter its shape in the PDS (i.e., for those cases for which the background rate is a significant fraction of the total observed rate). For our current work we decided to use all data since the determination of this effect is beyond the scope of our paper and we used  relatively small source extraction regions so the effects of background flaring should be relatively minor. However, we note that if one wants to constrain as accurately as possible the shape and the strength of the noise in the PDS, one has to investigate in more detail what the optimal size is for the source extraction region and the total amount of data that should be ignored due to high levels of background flaring.

\paragraph*{Effect of pile-up:} Although our sources are bright, they are not bright enough that significant amounts of pile-up are expected in our data. However, even in our data a certain degree of pile-up should be present (especially for the brightest sources in our example such us {\jsource} during its second {\it XMM-Newton} observation). The effect of this could in principle be studied by deselecting the inner RAWX column (or even multiple columns if the pile-up is really severe) from our source extraction region since in this column the observed source count rate is highest and the effects of pile-up the largest. However,  this would mean not using a large fraction of our data and this will reduce the significance of the PDS we can obtain. Again, since this effect depends strongly on the brightness of the source and our sources trace one order of magnitude in brightness, we will not investigate this effects further in our analysis. However, similarly to what has been found by \citet[][]{2004ApJ...601..439T} and \citet[][]{2013ApJ...766...89K} (who studied the effect of pile-up on PDS obtained with the ACIS CCD aboard {\it Chandra} and  with the {\it Swift}/XRT, respectively), we expect that pile-up during the {\it XMM-Newton} observations will decrease the exact level of the Poisson noise (which we correct for by estimating this level between 350 an 450 Hz before subtracting it from the PDS) and it will decrease the measured rms amplitudes of the noise components. Both studies found that the shape of the PDS did not change significantly so we expect that any possible pile-up in our data will not alter the observed shape of our {\it XMM-Newton} PDS as well.

\section{Energy dependency of the broad noise components}

For each source we created PDS in the following energy ranges: 0.5-10 keV, 0.5-1 keV, 1-2 keV, and 2-10 keV. For {\exo} we also created a PDS for the energy range 1-10 keV since below 1 keV the source was barely detected so the 0.5-10 keV PDS would be very close to the 1-10 keV PDS. We calculated the integrated noise between the frequencies $9.765625 \times 10^{-4}$ and 128 Hz ($2^{-10}$ - $2^7$ Hz) for all observations and for all energy ranges. The results are displayed in Table \ref{tab:energydependency}. We note that due to the uncertainties in the determination of the background, the rms amplitudes are lower limits although they should be close to the real value (see discussion in Appendix \ref{sec:timingtesting}), although for smaller energy selections this effects might increase, i.e., if the source count rate is low in this energy range (but studying this is beyond the scope of the current paper).

\newpage
\clearpage

\begin{table}
\caption{The energy dependence of the noise$^a$ in the PDS}
\begin{tabular}{llccccc}
\hline
Source                & ObsID           & 0.5-10 keV     &  0.5-1 keV      & 1-2 keV      & 2-10 keV      & 1-10 kev\\
                      &                 &  (\% rms)      &  (\% rms)       & (\% rms)     & (\% rms)      & (\% rms) \\
\hline 
\multicolumn{7}{c} {\underline {Sources identified as very hard by \citet{parikh2017}}}\\
{\jsource}            &  0741620101     & 33.5$\pm$0.2   & 36.8$\pm$0.7    & 33.0$\pm$0.3 & 33.0$\pm$0.2  &    -    \\
                      &  0741620201     & 18.2$\pm$0.1   & 14.4$\pm$0.5    & 18.7$\pm$0.2 & 21.6$\pm$0.2  &    -    \\
{\exo}                &  0744170201     & 38.3$\pm$0.3   & 48$\pm$12       & 36.3$\pm$0.9 & 39.0$\pm$0.3  & 38.3$\pm$0.3      \\
{\msource}            &  0701981401     & 102.0$\pm$1.9  & 91.6$\pm$2.4    & 95.8$\pm$1.9 & 109.9$\pm$2.0 &    -    \\
                      &  0701981501     & 91.9$\pm$1.2   & 78.1$\pm$1.6    & 84.6$\pm$1.2 & 100.0$\pm$1.2 &    -    \\
\hline
\multicolumn{7}{c}{\underline{Additional sources studied in the current work}}\\
{\igr}                &  0744840201   & 57.1$\pm$0.3 & 62.9$\pm$1.5 & 58.7$\pm$0.5 & 60.3$\pm$0.4 &    -    \\
{\ngcsource}          &  0790181401   & 26.2$\pm$0.7 & 16.7$\pm$5.8 & 24.8$\pm$1.7 & 34.4$\pm$1.3 &    -    \\
                      &  0790181501   & 30.9$\pm$0.7 & 23.8$\pm$4.4 & 28.8$\pm$1.6 & 39.4$\pm$1.4 &    -    \\
\hline
\multicolumn{7}{l} {$^a$ Noise integrated over the frequency range $9.765625 \times 10^{-4}$ - 128 Hz}\\
\end{tabular}
\label{tab:energydependency}
\end{table}

\label{lastpage}
\end{document}